\documentclass{aa}

\usepackage{txfonts}
\usepackage{graphicx}
\usepackage{array}
\usepackage{booktabs}
\usepackage[colorlinks=true,allcolors=blue]{hyperref}

\newcommand{\ffarcs}{\mbox{\ensuremath{.\!\!^{\prime\prime}}}}

\newcolumntype{L}[1]{>{\raggedright\let\newline\\\arraybackslash\hspace{0pt}}m{#1}}
\newcolumntype{C}[1]{>{\centering\let\newline\\\arraybackslash\hspace{0pt}}m{#1}}

\begin{document}

\title{MIRACLES: atmospheric characterization of directly imaged planets and substellar companions at 4--5~$\mu$m\thanks{Based on observations collected at the European Southern Observatory, Chile, ESO No. 085.C-0277(B), 090.C-0396(B), 095.C-0937(B), 199.C-0065(A), 0101.C-0588(A), and 0102.C-0649(A).}}

\subtitle{I. Photometric analysis of $\beta$~Pic~b, HIP~65426~b, PZ~Tel~B, and HD~206893~B}

\titlerunning{MIRACLES I. Photometric analysis of $\beta$~Pic~b, HIP~65426~b, PZ~Tel~B and HD~206893~B}

\author{
T.~Stolker\inst{1}
\and S.\,P.~Quanz\inst{1}\thanks{National Center of Competence in Research "PlanetS" (\url{http://nccr-planets.ch})}
\and K.\,O.~Todorov\inst{2}
\and J.~K\"{u}hn\inst{3,1}
\and P.~Molli\`{e}re\inst{4,5}
\and M.\,R.~Meyer\inst{6}
\and T.~Currie\inst{7,8}
\and S.~Daemgen\inst{1}
\and B.~Lavie\inst{9}
}

\institute{
Institute for Particle Physics and Astrophysics, ETH Zurich, Wolfgang-Pauli-Strasse 27, 8093 Zurich, Switzerland\\
\email{tomas.stolker@phys.ethz.ch}
\and Anton Pannekoek Institute for Astronomy, University of Amsterdam, Science Park 904, 1090 GE Amsterdam, The Netherlands
\and Center for Space and Habitability, University of Bern, Gesellschaftsstrasse 6, 3012, Bern, Switzerland
\and Leiden Observatory, Leiden University, Postbus 9513, 2300, RA Leiden, The Netherlands
\and Max-Planck-Institut f\"{u}r Astronomie, K\"{o}nigstuhl 17, 69117 Heidelberg, Germany
\and Department of Astronomy, University of Michigan, 1085 South University Avenue, Ann Arbor, MI 48109-1107, USA
\and NASA-Ames Research Center, Moffett Field, CA, USA
\and Subaru Telescope, National Astronomical Observatory of Japan, 650 North A`oh$\mathrm{\bar{o}}$k$\mathrm{\bar{u}}$ Place, Hilo, HI 96720, USA
\and Observatoire astronomique de l'Universit\'{e} de Gen\`{e}ve, 51 chemin des Maillettes, 1290 Versoix, Switzerland
}

\date{Received ?; accepted ?}

\abstract
{Directly imaged planets and substellar companions are key targets for the characterization of self-luminous atmospheres. Their photometric appearance at 4--5~$\mu$m is sensitive to the chemical composition and cloud content of their atmosphere.}
{We aim to systematically characterize the atmospheres of directly imaged low-mass companions at 4--5~$\mu$m. We want to homogeneously process the data, provide robust flux measurements, and compile a photometric library at thermal wavelengths of these mostly young, low-gravity objects. In this way, we want to find trends related to their spectral type and surface gravity by comparing with isolated brown dwarfs and predictions from atmospheric models.}
{We used the high-resolution, high-contrast capabilities of NACO at the Very Large Telescope (VLT) to directly image the companions of HIP~65426, PZ~Tel, and HD~206893 in the NB4.05 and/or $M'$ filters. For the same targets, and additionally $\beta$~Pic, we also analyzed six archival VLT/NACO datasets which were taken with the NB3.74, $L'$, NB4.05, and $M'$ filters. The data processing and photometric extraction of the companions was done with \texttt{PynPoint} while the \texttt{species} toolkit was used to further analyze and interpret the fluxes and colors.}
{We detect for the first time HIP~65426~b, PZ~Tel~B, and HD~206893~B in the NB4.05 filter, PZ~Tel~B and HD~206893~B in the $M'$ filter, and $\beta$~Pic~b in the NB3.74 filter. We provide calibrated magnitudes and fluxes with a careful analysis of the error budget, both for the new and archival datasets. The $L'$~--~NB4.05 and $L'$~--~$M'$ colors of the studied sample are all red while the NB4.05~--~$M'$ color is blue for $\beta$~Pic~b, gray for PZ~Tel~B, and red for HIP~65426~b and HD~206893~B (although typically with low significance). The absolute NB4.05 and $M'$ fluxes of our sample are all larger than those of field dwarfs with similar spectral types. Finally, the surface gravity of $\beta$~Pic~b has been constrained to $\log{g} = 4.17_{-0.13}^{+0.10}$~dex from its photometry and dynamical mass.}
{A red color at 3--4~$\mu$m and a blue color at 4--5~$\mu$m might be (partially) caused by H$_2$O and CO absorption, respectively, which are expected to be the most dominant gaseous opacities in hot ($T_\mathrm{eff} \gtrsim 1300$~K) atmospheres. The red characteristics of $\beta$~Pic~b, HIP~65426~b, and HD~206893~B at 3--5$\mu$m, as well as their higher fluxes in NB4.05 and $M'$ compared to field dwarfs, indicate that cloud densities are enhanced close to the photosphere as a result of their low surface gravity.}

\keywords{Stars: individual: $\beta$~Pictoris, HIP~65426, PZ~Tel, HD~206893 -- Planets and satellites: atmospheres -- Methods: data analysis -- Techniques: high angular resolution, image processing}

\maketitle

\section{Introduction}\label{sec:introduction}

\begin{table*}
\caption{Target information.}
\label{table:targets}
\centering
\bgroup
\def\arraystretch{1.25}
\begin{tabular}{L{2cm} C{2cm} C{2cm} C{1.3cm} C{2cm} C{2cm} C{3.2cm}}
\hline\hline
Target & Spectral type & Distance\tablefootmark{a} & Age & $L'$ / $W1$\tablefootmark{b} & $M'$ / $W2$\tablefootmark{b} & References \\
 & & (pc) & (Myr) & (mag) & (mag) & \\
\hline
$\beta$~Pic & A6V & $19.75 \pm 0.13$ & $22 \pm 6$ & $3.454 \pm 0.003$ & $3.458 \pm 0.009$ & (1), (2), (3), (4) \\
HIP~65426 & A2V & $109.21 \pm 0.75$ & $14 \pm 4$ & $6.761 \pm 0.038$ & $6.798 \pm 0.019$ & (3), (5), (6), (7) \\
PZ~Tel & G9IV & $47.13 \pm 0.13$ & $24 \pm 3$ & $6.257 \pm 0.049$ & $6.285 \pm 0.022$ & (3), (7), (8), (9) \\
HD~206893 & F5V & $40.81 \pm 0.11$ & $250_{-200}^{+450}$ & $5.528 \pm 0.066$ & $5.437 \pm 0.028$ & (1), (3), (7), (10) \\
\hline
\end{tabular}
\egroup
\tablefoot{\\
\tablefoottext{a}{Distances are calculated from the \emph{Gaia} DR2 parallaxes.}\\
\tablefoottext{b}{$L'$ and $M'$ magnitudes in the ESO system for $\beta$~Pic~b and WISE $W1$ and $W2$ magnitudes for all other targets.}\\
\textbf{References.} (1) \citet{gray2006}, (2) \citet{shkolnik2017}, (3) \citet{gaia2018}, (4) \citet{bouchet1991}, (5) \citet{houk1978}, (6) \citet{chauvin2017}, (7) \citet{cutri2012}, (8) \citet{jenkins2012}, (9) \citet{torres2006}, (10) \citet{delorme2017}.
}
\end{table*}

The population of directly imaged planetary and substellar companions provides an important window onto the formation and evolution of low-mass objects on long-period orbits \citep[e.g.,][]{bowler2016}. Large-scale surveys have been ongoing for more than a decade, yielding a dozen exoplanet detections \citep[e.g.,][]{marois2008,lagrange2009,macintosh2015,chauvin2017} and constraints on their demographics beyond $\sim$10~au \citep[e.g.,][]{stone2018,nielsen2019}. Sensitivity limits continue to increase thanks to dedicated high-contrast imaging instruments \citep[e.g.,][]{macintosh2008,jovanovic2015,beuzit2019} and observing strategies \citep[e.g.,][]{kaufli2018}, as well as developments in image processing techniques and detection algorithms \citep[e.g.,][]{gomez-gonzalez2018,flasseur2018}. As a result, detailed spectrophotometric measurements have been carried out at near-infrared (NIR) wavelengths to constrain some of the physical and chemical atmospheric properties \citep[e.g..,][]{derosa2016,samland2017}. At the same time, multi-epoch observations are starting to place constraints on the orbital architecture of these objects \citep[e.g.,][]{wang2018}, sometimes in tandem with stellar radial velocity and/or astrometry constraints \citep[e.g.,][]{bonnefoy2018,dupuy2019}.

Atmospheric studies of directly imaged planets have benefitted from observations of isolated brown dwarfs \citep[e.g.,][]{golimowski2004,cushing2008,stephens2009,leggett2010,yamamura2010} and extended modeling efforts of their evolution and atmospheric processes \citep[e.g.,][]{ackerman2001,saumon2008,allard2012,baraffe2015}. Without the bright glare of a central star, these objects have been detected in the solar neighborhood \citep{luhman2013} down to temperatures of only a few hundred Kelvin \citep{skemer2016b,morley2018}. Gas giant planets and brown dwarfs are expected to share some of their main atmospheric characteristics, such as temperature, radius, and composition, as well as the physics and chemistry that govern their atmospheres \citep[e.g.,][]{marley2015}. As a result, the directly imaged planets share a similar spectral sequence to that of their older counterparts. However, for a given temperature, their mass and related surface gravity are lower due to a difference in age, which leads to noticeable differences in their appearance. For example, the NIR colors of young low-mass companions appear redder while their absolute fluxes are lower \citep[e.g.,][]{metchev2006,marois2008,barman2011}. These characteristics can be attributed to thicker and/or vertically more extended clouds in a low-gravity environment \citep[e.g.,][]{currie2011}, as well as the typical sizes of the condensed dust grains \citep[e.g.,][]{burrows2006,marley2012}. The surface gravity may also impact the chemical abundances because enhanced vertical mixing can lead to a nonequilibrium balance of CO and CH$_4$ \citep[e.g.,][]{yamamura2010,zahnle2014,moses2016}.

The advent of extreme adaptive optics (AO) assisted high-contrast imaging instruments has provided detailed insight into the NIR (1--2.4~$\mu$m) photometric and spectral characteristics of the family of directly imaged low-mass companions \citep[e.g.,][]{vigan2016,chilcote2017,greenbaum2018}, but their mid-infrared (MIR; 3--5~$\mu$m) characteristics remain more sparsely sampled \citep[e.g.,][]{galicher2011,bailey2013,skemer2014,stone2016,skemer2016a,cheetham2019}. While observations at these wavelengths are hampered by the bright thermal background emission, the flux contrast with the central star is more favorable and an increasing number of photons is emitted at MIR wavelengths along the spectral sequence towards lower mass and cooler objects \citep[see Fig.~1 from][]{skemer2014}. Most importantly, complementary information about the chemical composition and cloud configuration are to be expected from photometry at 3--5~$\mu$m \citep[e.g.,][]{geisler2008,galicher2011,currie2014}. The \emph{James Webb Space Telescope (JWST)} will fully exploit the MIR regime \citep{boccaletti2015,danielski2018} -- in particular for companions with moderate contrast at wide separations -- while ground-based telescopes provide direct access to the atmospheres of more close-in planets \citep{mawet2016,kenworthy2018}.

The MIRACLES (Mid-InfraRed Atmospheric Characterization of Long-period Exoplanets and Substellar companions) survey was designed for photometric characterization of directly imaged planetary and substellar companions at 4--5~$\mu$m. It was carried out with the NAOS-CONICA (NACO) AO system and infrared camera \citep{lenzen2003,rousset2003} that was mounted on the Unit Telescope 1 (UT1) of the Very Large Telescope (VLT) at the Paranal Observatory in Chile. The sample consists of 15 targets that were observed with the Brackett-$\alpha$ and/or $M'$ filters. We aim to homogeneously process and analyze the data in order to obtain robust fluxes and colors of the companions. For consistency, we also reprocess archival data in the 4--5~$\mu$m range where such data are available, as well as archival $L'$ band data to cover a second spectral window. The current paper presents the first results as well as extended details on the data processing and calibration. Therefore, only a subset of the companions ($\beta$~Pic~b, HIP~65426~b, PZ~Tel~B, and HD~206893~B) is analyzed as a demonstration while a more in-depth analysis of the full sample will follow in a future work. Some of the main stellar characteristics that are relevant for this study are listed in Table~\ref{table:targets}.

In Sect.~\ref{sec:observations}, we report the high-contrast imaging observations with VLT/NACO and the archival datasets that have been (re)processed. In Sect.~\ref{sec:data_reduction}, we describe the data processing with \texttt{PynPoint}, present the companion detections, and outline the procedure for the photometric extraction, calibration, and uncertainty estimation. In Sect.~\ref{sec:results}, we present the photometry and colors, and compare them with results for field and low-gravity brown dwarfs, other directly imaged companions, and synthetic model photometry. We also carry out a dedicated analysis of $\beta$~Pic~b and present the detection limits derived from our data. In Sect.~\ref{sec:discussion}, we discuss our measurements and provide an outlook on future opportunities.

\begin{table*}
\caption{Observation details on the new and archival data.}
\label{table:observations}
\centering
\bgroup
\def\arraystretch{1.25}
\begin{tabular}{L{1.7cm} L{1.4cm} L{1.9cm} C{0.9cm} C{1.1cm} C{0.8cm} C{1.5cm} C{1.5cm} C{1.2cm} C{1.7cm}}
\hline\hline
Target & Filter & UT date & DIT\tablefootmark{a} & $N_\mathrm{images}$ & $\Delta\pi$\tablefootmark{b} & Airmass & Seeing\tablefootmark{c} & $\tau_0$\tablefootmark{d} & FWHM\tablefootmark{e} \\
 & & & (s) & & (deg) & & (arcsec) & (ms) & (mas) \\
\hline
\multicolumn{2}{l}{\emph{New data}}\\
HIP~65426 & NB4.05 & 2018 Jun 7 & 1.0 & 11520 & 99.5 & 1.12--1.22 & 1.52(3.01) & 3.5(0.6) & $112.8 \pm 0.9$ \\
PZ~Tel & NB4.05 & 2018 Jun 8 & 1.0 & 1160 & 35.3 & 1.11--1.19 & 0.82(0.06) & 4.2(0.5) & $112.5 \pm 1.1$ \\
PZ~Tel & $M'$ & 2018 Jun 8 & 0.04 & 40000 & 25.1 & 1.13--1.21 & 0.85(0.07) & 4.9(1.0) & $125.9 \pm 2.2$ \\
HD~206893 & NB4.05 & 2018 Oct 24 & 0.9 & 3752 & 68.5 & 1.02--1.04 & 1.02(0.10) & 3.1(0.6) & $111.6 \pm 0.9$ \\
HD~206893 & $M'$ & 2018 Jun 8 & 0.04 & 115000 & 85.5 & 1.02--1.08 & 0.89(0.09) & 4.5(0.6) & $121.7 \pm 1.1$ \\
\hline
\multicolumn{2}{l}{\emph{Archival data}}\\
$\beta$~Pic & NB3.74\tablefootmark{f} & 2012 Dec 18 & 0.1 & 12000 & 43.1 & 1.12--1.18 & 0.88(0.05) & 4.3(0.6) & $108.2 \pm 1.1$  \\
$\beta$~Pic & NB4.05 & 2012 Dec 16 & 0.075 & 25600 & 57 & 1.12--1.15 & 0.74(0.02) & 5.5(1.0) & $113.4 \pm 1.1$ \\
HIP~65426 & $L'$ & 2017 May 18 & 0.2 & 30224 & 86.9 & 1.12--1.33 & 0.90(0.11) & 3.8(0.5) & $107.3 \pm 1.0$ \\
HIP~65426 & $M'$ & 2017 May 20 & 0.05 & 72900 & 73.3 & 1.12--1.35 & 0.64(0.04) & 5.2(1.2) & $135.4 \pm 0.9$ \\
PZ~Tel & $L'$ & 2010 Sep 26 & 0.3 & 14400 & 42.4 & 1.11--1.21 & 3.60(2.54) & 0.9(0.1) & $99.2 \pm 1.7$ \\
HD~206893 & $L'$ & 2016 Aug 9 & 0.3 & 18000 & 96.7 & 1.04--1.09 & 0.86(0.09) & 5.7(1.2) & $117.0 \pm 1.8$ \\
\hline
\end{tabular}
\egroup
\tablefoot{The upper part of the table lists the observed targets and the lower part the archival datasets that were analyzed.\\
\tablefoottext{a}{The total integration time is given by the product of detector integration time (DIT) and the total number of images ($N_\mathrm{images}$).}\\
\tablefoottext{b}{Total parallactic rotation.}\\
\tablefoottext{c}{Mean and standard deviation of the seeing as measured by the differential image motion monitor (DIMM) at 0.5~$\mu$m.}\\
\tablefoottext{d}{Mean and standard deviation of the coherence time.}\\
\tablefoottext{e}{Full width at half maximum of the unsaturated point spread function (see main text for details).}\\
\tablefoottext{f}{Archival data which has not yet been published.}\\
}
\end{table*}

\section{Observations and archival data}\label{sec:observations}

\subsection{High-contrast imaging with VLT/NACO at 4--5~$\mu$m}\label{sec:naco_imaging}

The data were acquired with VLT/NACO at the Paranal Observatory in Chile (ESO program IDs: 0101.C-0588(A) and 0102.C-0649(A)). We used the narrowband Brackett-$\alpha$ (NB4.05) filter ($\lambda_0 = 4.05$~$\mu$m, $\Delta\lambda = 0.02$~$\mu$m) and the broadband $M'$ filter ($\lambda_0 = 4.78$~$\mu$m, $\Delta\lambda = 0.59$~$\mu$m) to probe two complementary parts of the 4--5~$\mu$m regime. The upper part of Table~\ref{table:observations} provides an overview of the observed targets and their filters.

We used the pupil-stabilized mode of the instrument to detect the off-axis companions during post-processing with angular differential imaging (ADI). The targets were dithered between the top left and bottom right quadrant of the detector. The bottom left quadrant was excluded due to the persistent striping while the top right quadrant was avoided out of precaution since it was occasionally affected by an excess of striped detector noise. The number of integrations per dither position and readout corresponded to an effective exposure time of $\sim$1~min such that slowly evolving variations in the thermal background emission could be sampled. We used the \emph{Uncorr} readout mode, which resets and reads the detector array once for each integration and is therefore most suitable for observations with a high background flux. The pixel scale for the L27 camera has not been recalibrated so we assume a value of 27.1~mas per pixel.

At thermal wavelengths ($M'$ band in particular), the detector integration time (DIT) is typically limited by the brightness of the background emission except for bright targets such as $\beta$~Pic. Therefore, the observations did not require the use of a coronagraph (which was also not available for the $M'$ filter). The DIT that could be used in $M'$ was 40~ms while integrations of 0.9 or 1.0~s were used with the NB4.05 filter (see Table~\ref{table:observations}). Given the short exposure time with the $M'$ filter, we had windowed the detector to a field of view of 256~pixels in order to prevent frame loss and to limit the amount of overhead time. The data were obtained in cube mode which means that each individual exposure was stored such that a careful frame selection was possible.

The flux of the companions was measured relative to their central star (see Sect.~\ref{sec:relative_calibration}) which requires unsaturated exposures of the stellar point spread function (PSF). The peak flux in $M'$ remained at least 50\% below the full well depth of the detector, which corresponds to 28000~ADU with the \emph{HighBackground} detector mode (the setting of the bias voltage of the array). For the NB4.05 data, the peak flux had values up to two-thirds of the full well depth (i.e., 15000~ADU with the \emph{HighDynamic} detector mode). Therefore, we obtained additional exposures with a smaller DIT for the flux calibration to ensure a robust sampling of the stellar flux: specifically, 480 exposures of 0.5--0.6~s for HIP~65426, 720 exposures of 0.5~s for PZ~Tel, and 600 exposures of 0.4~s for HD~206893. After processing of the images (see Sect.~\ref{sec:data_processing}), the stellar PSF was fitted with a 2D Moffat profile to determine the angular resolution of the data. The best-fit values of the full width at half maximum (FWHM) of each PSF are listed in Table~\ref{table:observations}, for which the averages of the major and minor axes were used.

\begin{figure*}
\centering
\includegraphics[width=\linewidth]{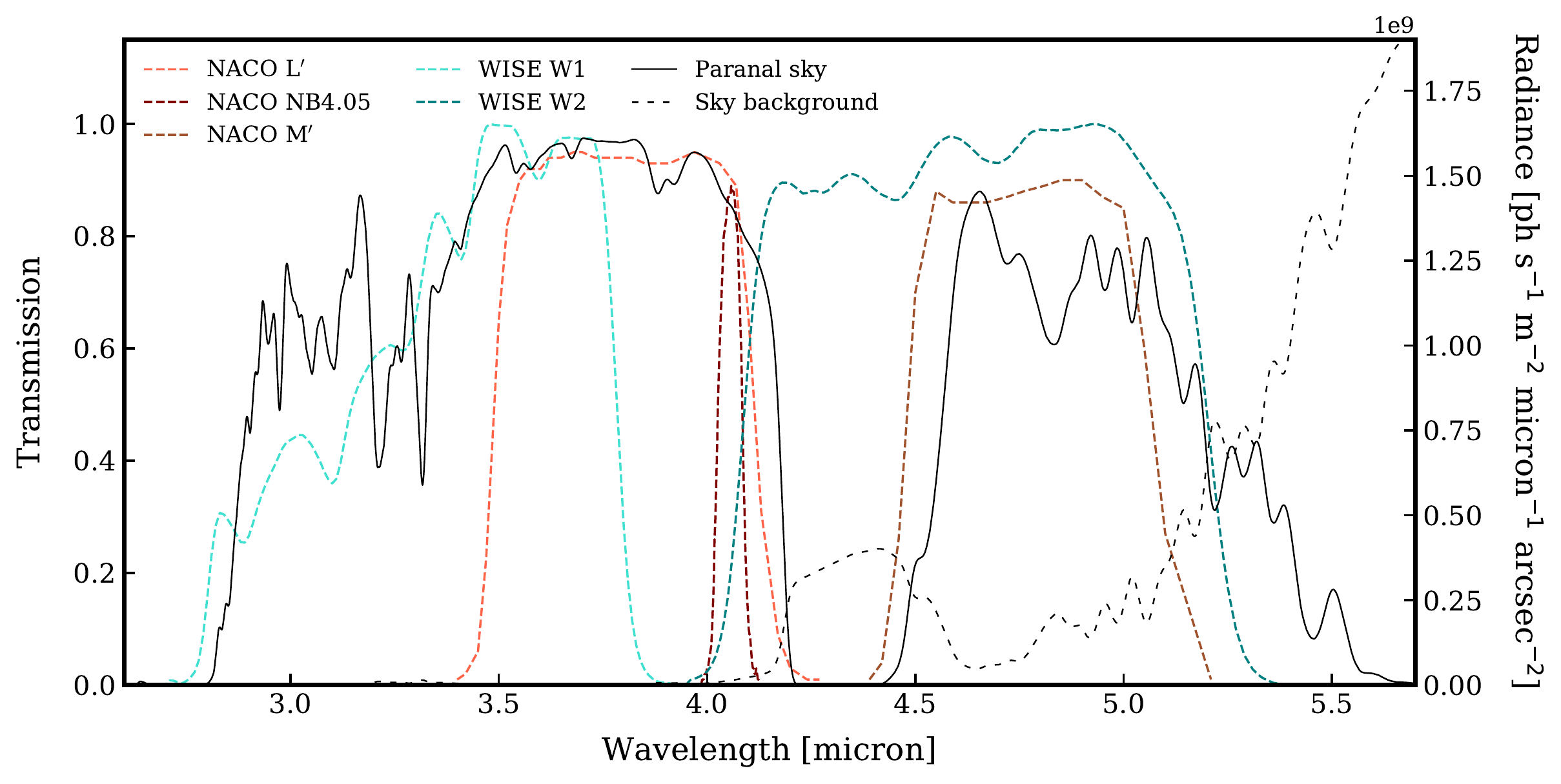}
\caption{Transmission profiles of the NACO $L'$, NB4.05, and $M'$ filters and the WISE $W1$ and $W2$ filters (\emph{colored dotted lines}). The telluric transmission at the Paranal Observatory is also shown (\emph{black solid line}). The right axis shows the sky background emission (\emph{black dashed line}) which steeply increases beyond $\sim$4.5~$\mu$m.\label{fig:transmission}}
\end{figure*}

The observing conditions were overall stable on the night of 2018~June~7 with a mean seeing of about 0\ffarcs8--0\ffarcs9. The seeing made a temporary sharp increase during the observations of HIP~65426 but remained below 1\ffarcs0 for most of the sequence. Such low-quality images were removed during the subsequent frame selection (see Sect.~\ref{sec:data_processing}). The conditions were poorer and more variable during the observations of HD~206893 in NB4.05 with the seeing fluctuating between 0\ffarcs9 and 1\ffarcs3. The mean coherence time was typically in the range of 3--5~ms (see Table~\ref{table:observations}). Generally, observations at 4--5~$\mu$m yield higher Strehl ratios than at shorter wavelengths because for a given phase aberration the relative impact on a PSF in the $M'$ band is significantly smaller than in the optical. Consequently, speckle noise is typically less important for companion detections at 4--5~$\mu$m.

The transmission profiles of the considered NACO filters are shown in Fig.~\ref{fig:transmission} in comparison with the WISE $W1$ and $W2$ filters, which cover the $L'$ and $M'$ bands, respectively. The WISE photometry is used in Sect.~\ref{sec:absolute_calibration} to calibrate the extracted fluxes of the companions. The figure also shows the transmission at Cerro Paranal, which was computed with the \emph{SkyCalc} interface \citep{noll2012} for an airmass of 1.0 and a precipitable water vapor level of 2.5~mm (approximately the median value at the observatory). In addition to the limited telluric transmission in $M'$ ($\sim$70\%), the steep increase of the sky background emission across this band is also visible. The majority of the sky radiance at wavelengths longer than 4~$\mu$m is caused by molecular emission in the lower part of the Earth's atmosphere.

\subsection{Complementary and reprocessed archival data}\label{sec:archival_data}

The five new NACO datasets that were obtained with the NB4.05 and $M'$ filters were complemented with several archival datasets. Here we selected NB4.05 and/or $M'$ datasets in case these were already available, $L'$ band data for all targets, and a dataset of $\beta$~Pic~b taken with the narrowband NB3.74 filter (which overlaps the central part of the $L'$ bandpass). We chose to reprocess these archival datasets in order to homogeneously analyze them with the newly obtained data such that the photometric extraction, calibration, and uncertainty estimation is done in a consistent manner. By including $L'$ data, we obtain additional color information in the 3--4~$\mu$m spectral region, which is complementary to the 4--5~$\mu$m range.

The archival data that were used in this study are listed in the lower part of Table~\ref{table:observations}. Here we briefly summarize a few of the relevant data characteristics and observing conditions. For $\beta$~Pic, we analyzed data that were obtained with the NB3.74 and NB4.05 filters (ESO program ID: 090.C-0396(B)) of which the NB4.05 data were published by \citet{currie2013}. The photometry in these filters is combined in the dedicated analysis of this object in Sect.~\ref{sec:beta_pic}. With integration times of 0.1~s and 0.075~s, the stellar flux had remained within the linear detector regime for both filters so we could use the full sequence of images for the photometric calibration.

For HIP~65426, we reprocessed the $L'$ and $M'$ band data (ESO program ID: 199.C-0065(A)) that were recently published by \citet{cheetham2019}. Separate exposures with a DIT of 0.1~s (4~min in total) had been taken in $L'$ for the flux calibration while the DIT of the $M'$ data was sufficiently low such that the images themselves could be used as PSF template. The conditions were variable during the $L'$ observations with a seeing in the range of 0\ffarcs8--1\ffarcs1, which may impact the precision of the flux calibration. The $M'$ data on the other hand were obtained in good observing conditions.

For PZ~Tel, we complemented our study with archival $L'$ data (ESO program ID: 085.C-0277(B)) that were published by \citet{beust2016}. The data were taken in highly variable conditions with a mean seeing of 3\ffarcs6 and values below 1\ffarcs0 for only about 20\% of the observing sequence. Nonetheless, the star had remained visible with sufficient AO correction but the flux of PZ~Tel varied up to $\sim$40\%. Additional exposures were taken with a smaller DIT of 0.2~s (8~min in total) and a neutral density filter (\texttt{ND\_long}) in the optical path.

Finally, we reprocessed the $L'$ band data (ESO program ID: 095.C-0937(B)) of HD~206893 that were presented by \citet{milli2017}. The data were taken with the annular groove phase mask coronagraph with six additional off-axis exposures of the star with a DIT of 0.1~s for calibration purpose. In this case, the background emission was sampled by nodding the field of view away from the target every $\sim$10~min.

\section{Data reduction, detection, and calibration}\label{sec:data_reduction}

\subsection{Data processing with PynPoint}\label{sec:data_processing}

The data reduction was done with \texttt{PynPoint}\footnote{\url{https://github.com/PynPoint/PynPoint}}, which is an end-to-end pipeline for processing and analysis of high-contrast imaging data \citep{amara2012,stolker2019}. In Sect.~\ref{sec:photometry}, we also use its functionalities for the extraction of the relative flux and uncertainty of the companions, and in Sect.~\ref{sec:limits} for estimating the detection limits from our data. All results were obtained with version 0.8.1, which is currently the latest release \footnote{\url{https://pypi.org/project/pynpoint/}}.

After reading the raw FITS files into the database, we applied basic preprocessing and calibration procedures. The thermal background emission, dark current, and detector bias were subtracted with the mean of the adjacent data cubes in which the star was located at a different dither position. Remaining bad pixels were corrected by selecting 5$\sigma$ outliers within a $9 \times 9$~pixel filter and replacing them with the mean of the neighboring values. The parallactic angle associated with each individual exposure was precisely calculated from the relevant header information.

Frames were registered by first cropping a subregion around the brightest pixel and subsequently, for a relative alignment, cross-correlating each image with ten randomly selected references images and shifting to the mean offset with the reference frames. Finally, for an absolute centering,the mean of the image stack was fitted with a 2D Moffat function and a constant shift was applied to each image. After the first registration step, we also applied a frame selection by measuring the integrated flux within an aperture (1~FWHM in radius) centered at the approximate position of the star and removing poor-quality frames by sigma clipping frames of which the flux deviated by more than 1--2$\sigma$ from the median. Typically, about 5--20\% of the images were removed with the frame selection except for the NB3.74 dataset of $\beta$~Pic and the $L'$ dataset of PZ~Tel. For the first, we removed 31\% of the data due to the degrading observing conditions at the end of the sequence. For the latter, we only used 11\% of the data, which were selected from the end of the sequence during which the photometry remained approximately stable and after which the flux exposures were taken.

Subsequently, we stacked subsets of images such that the final image stack contained approximately 500 images. This is required for the Markov chain Monte Carlo (MCMC) analysis of the  contrast and position of the companions (see Sect.~\ref{sec:photometry}), which otherwise would be too computationally expensive. However, it also typically enhances the signal-to-noise ratio (S/N) for the companion detections in the $L'$ and $M'$ filters \citep[see e.g.,][]{meshkat2014,quanz2015b}. Images were also cropped before running the PSF subtraction and the MCMC analysis, but a larger field of view was used for the calculation of the detection limits (see Sect.~\ref{sec:limits}).

The PSF subtraction was done with an implementation of full-frame principal component analysis \citep[PCA;][]{amara2012,soummer2012}. First, pixel values were masked at radii larger than the image size and at separations within typically 1~FWHM of the image center. Second, the stack of images were decomposed into a lower-dimensional basis set of orthogonal images by applying a singular value decomposition (SVD). Each image was then projected onto the basis of principal components (PCs) and the model was subtracted from the image itself to remove the quasi-static PSF and speckle noise from the star. We varied the number of PCs that were used for the PSF subtraction in the range of 1--50. Finally, all images were derotated towards a common field orientation and median-combined.

\subsection{Companion detections}\label{sec:detections}

\begin{figure*}
\centering
\includegraphics[width=\linewidth]{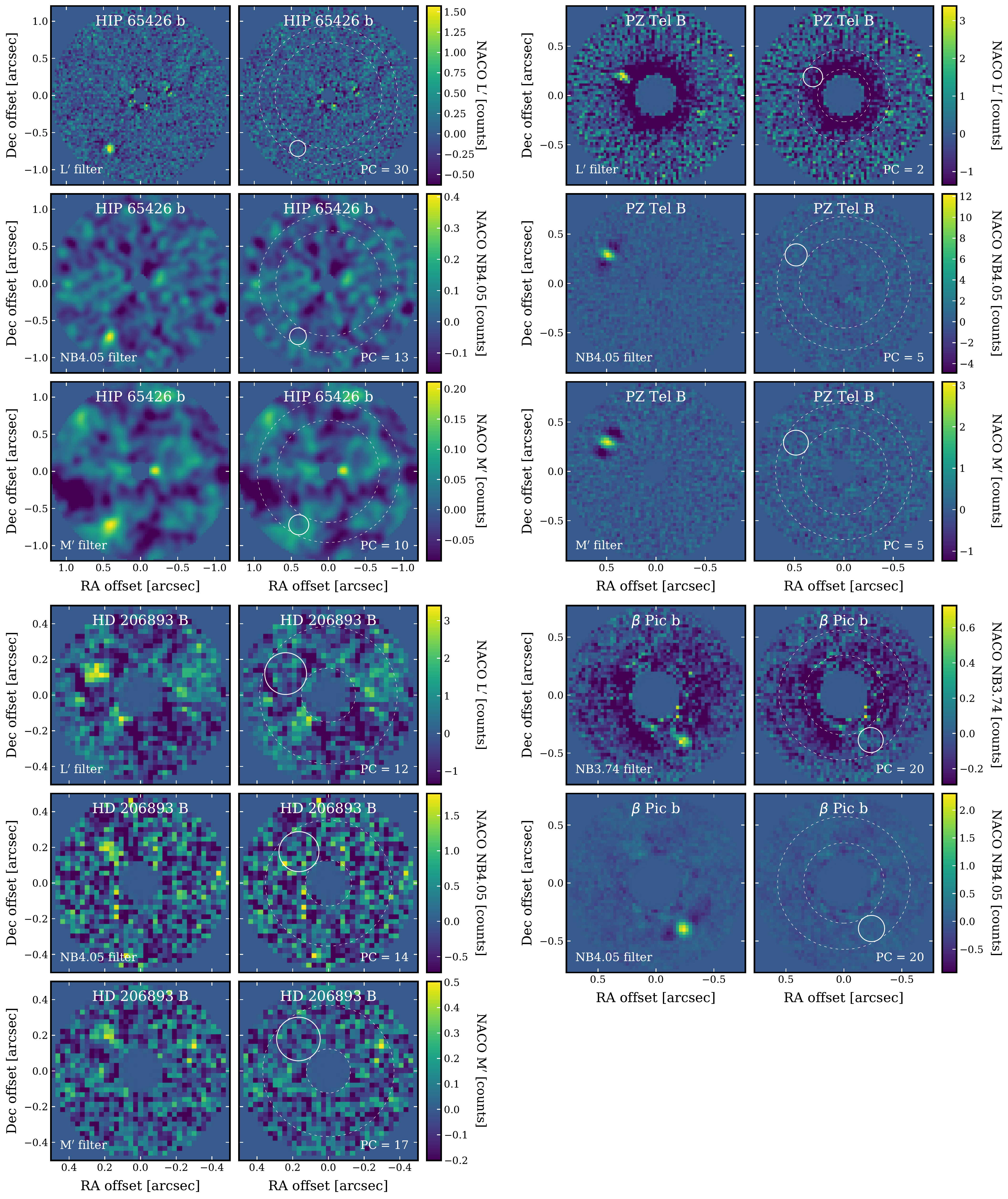}
\caption{Median-combined residuals of the PSF subtraction for HIP~65426 (\emph{top left}), PZ~Tel (\emph{top right}), HD~206893 (\emph{bottom left}), and $\beta$~Pic (\emph{bottom right}). Images are ordered for each target from top to bottom by the central wavelength of the filters. For each object, the left column shows the regular residuals of the PSF subtraction while the right column shows the residuals with the best-fit negative PSF template injected. The color scale is set by the maximum pixel value of each object and filter combination and adjusted by a factor -0.3 of the peak flux for the lower limit. The pixel values within the \emph{solid circles} (at the positions of the companions) were minimized for the flux estimation while the noise was estimated from the area indicated by the \emph{dashed circles} (see Sect.~\ref{sec:relative_calibration} for details). North and east are in upward and leftward direction, respectively, in all images.\label{fig:images}}
\end{figure*}

The residuals of the PSF subtraction are presented in the first and third columns of Fig.~\ref{fig:images}. We calculated the S/N for each number of PCs with the two-sample \emph{t}-test, which includes a correction term for the small sample statistics \citep{mawet2014}:
\begin{equation}
\mathrm{S/N} = \frac{\bar{x}_1-\bar{x}_2}{s_2\sqrt{1+\frac{1}{n_2}}},
\end{equation}
where $\bar{x}_1$ is the flux at the position of the companion, $\bar{x}_2$ the average flux within the remaining nonoverlapping apertures at the same separation, $s_2$ the empirical standard deviation, and $n_2$ the number of reference apertures. We chose a conservative aperture diameter of 1~FWHM and excluded the apertures directly adjacent to the companion aperture as they contained self-subtraction artifacts. The S/N was then optimized with the position of the companion aperture as a free parameter. The final flux calibration was done for a fixed number of PCs, which was chosen by maximizing the S/N and limiting the amount of variation in the retrieved contrast and position values (see Sect.~\ref{sec:relative_calibration}).

While $\beta$~Pic~b and PZ~Tel~B are bright targets at 4--5~$\mu$m and detected with high S/N (maximum values in the range of 10--35), HIP~65426~b is significantly fainter and only reaches moderately above the background limit in the NB4.05 and $M'$ filters (maximum S/N of 8.1 and 5.5, respectively). For these filters, we smoothed the images of HIP~65426~b with a Gaussian filter of similar FWHM to the angular resolution in order to lower pixel-to-pixel variations and enhance the planet detection (but the original images were used for the photometric extraction). HD~206893~B is brighter than HIP~65426~b but the residual speckle noise at its small angular separation ($\sim$2--3$\lambda/D$ at the observed wavelengths) limited the S/N to values in the range of 5--8. Nonetheless, this companion can also be identified after subtracting a sufficient number of PCs from the data. Although PZ~Tel~B was detected in $L'$ with an S/N of 51 when the full dataset was used, we only considered a small subset of the images to ensure a robust flux calibration (see Sect.~\ref{sec:observations}) which resulted in a reduced S/N of 12.8.

\begin{figure*}
\centering
\includegraphics[width=\linewidth]{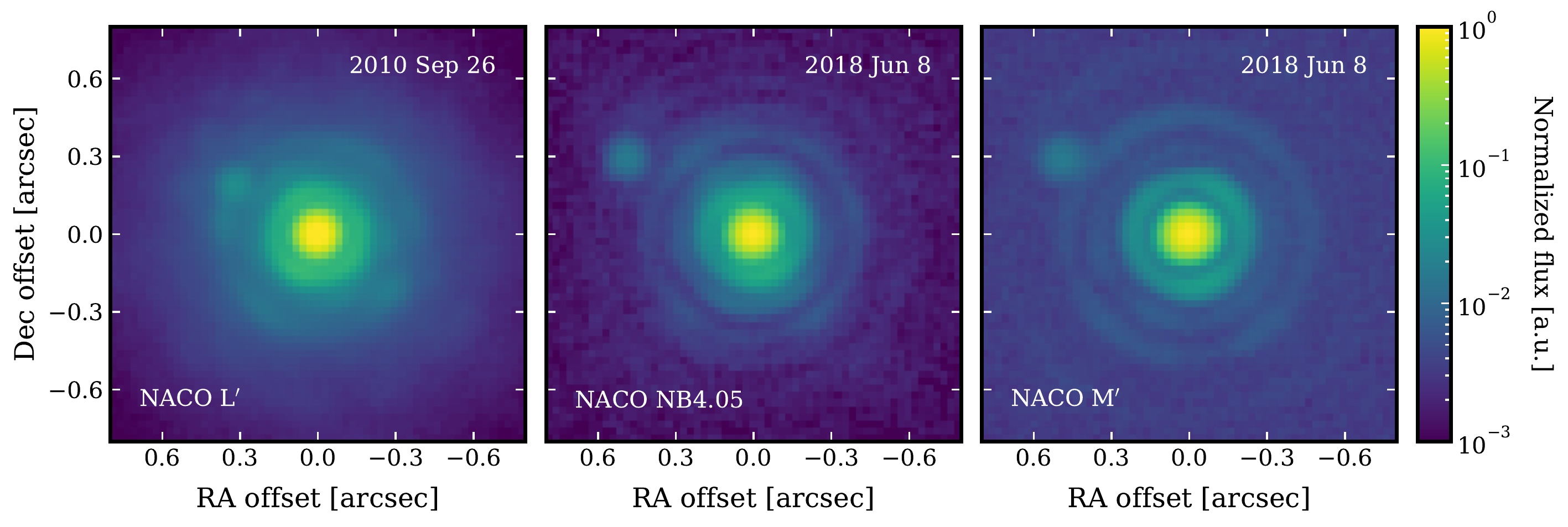}
\caption{Derotated and median-combined images of PZ~Tel in the $L'$ (\emph{left}), NB4.05 (\emph{center}), and $M'$ (\emph{right}) filters. The $\sim$8~yr baseline with respect to the archival $L'$ data reveals clear orbital motion. The colors are shown on a logarithmic scale which has been normalized to the peak flux. North and east are in upward and leftward direction, respectively. \label{fig:pztel}}
\end{figure*}

In addition to the PSF subtraction residuals, we display in Fig.~\ref{fig:pztel} the derotated and median-combined images of PZ~Tel in all three filters. The $L'$ band data were obtained $\sim$8~years before the NB4.05 and $M'$ data, and therefore the companion is detected in $L'$ at a
separation that is smaller by 190~mas. Although the peak flux of the PSF is slightly saturated at the shortest wavelengths, it can be seen from the images that PZ~Tel~B is approximately two orders of magnitude fainter than PZ~Tel~A in all filters. The image quality in $L'$ is affected by the poor observing conditions while the NB4.05 and $M'$ images appear of better quality with a well discernible Airy pattern.

\subsection{Photometric extraction, calibration, and error estimation}\label{sec:photometry}

\subsubsection{Relative calibration by negative PSF injection}\label{sec:relative_calibration}

We require accurate and robust photometry of our companions to characterize their atmospheres. Photometric and astrometric measurements of directly imaged objects are challenging due to self-subtraction effects that are intrinsic to post-processing algorithms such as PCA. One way to overcome this is problem is by iteratively injecting negative copies of the unsaturated stellar PSF and minimizing the residuals at the position of the companion after the PSF subtraction. For this procedure, we masked the PSF beyond a radius of 2--4~FWHM, depending on its brightness relative to the background flux.

The contrast and position were first computed as a function of the number of PCs by minimizing the flux residuals with a downhill simplex method. Similarly to \citet{stolker2019}, we then used Bayesian inference with MCMC to sample the posterior distributions of the separation, position angle, and flux contrast of the companions in order to determine the most probable values and their uncertainties. For the MCMC, the number of PCs was fixed (i.e., the values indicated in Fig.~\ref{fig:images}) by selecting a value for which both the S/N was high and the retrieved contrast and position values showed minimal variation.

The log-likelihood function for the MCMC was defined as
\begin{equation}\label{eq:likelihood}
\log \mathcal{L} \propto -\frac{1}{2} \sum_{i,j}^N \left(\frac{I_{ij}}{\sigma_\mathrm{pix}}\right)^2,
\end{equation}
where $i$ and $j$ are the pixel indices, $N$ is the total number of pixels encircled by the aperture, $I_{ij}$ is the pixel value at position $ij$, and $\sigma_\mathrm{pix}$ is the (constant) noise level associated with the pixels. The pixel values that are minimized have been selected within a circular aperture with a diameter of 2~FWHM at the position of the companion (see Fig.~\ref{fig:images}). The noise on the other hand is calculated by derotating the residuals of the PSF subtraction in the opposite direction, median-combining the images, and computing the standard deviation of all pixels within an annulus covering the separation range of the circular aperture.

When defining the likelihood function, we make several assumptions. First, we assume that the expected value is zero when the injected negative PSF has fully removed the companion signal. Second, the pixel values are considered as independent measurements which may not be strictly true due to potential spatial correlations in the noise residuals after the PSF subtraction. Third, the reference pixels within the annulus are assumed to follow a Gaussian distribution. This is in contrast to the approach followed in \citet{stolker2019}, where the noise associated with each pixel was assumed to follow a Poisson distribution \citep[see also][]{wertz2017}.

\begin{figure*}
\centering
\includegraphics[width=\linewidth]{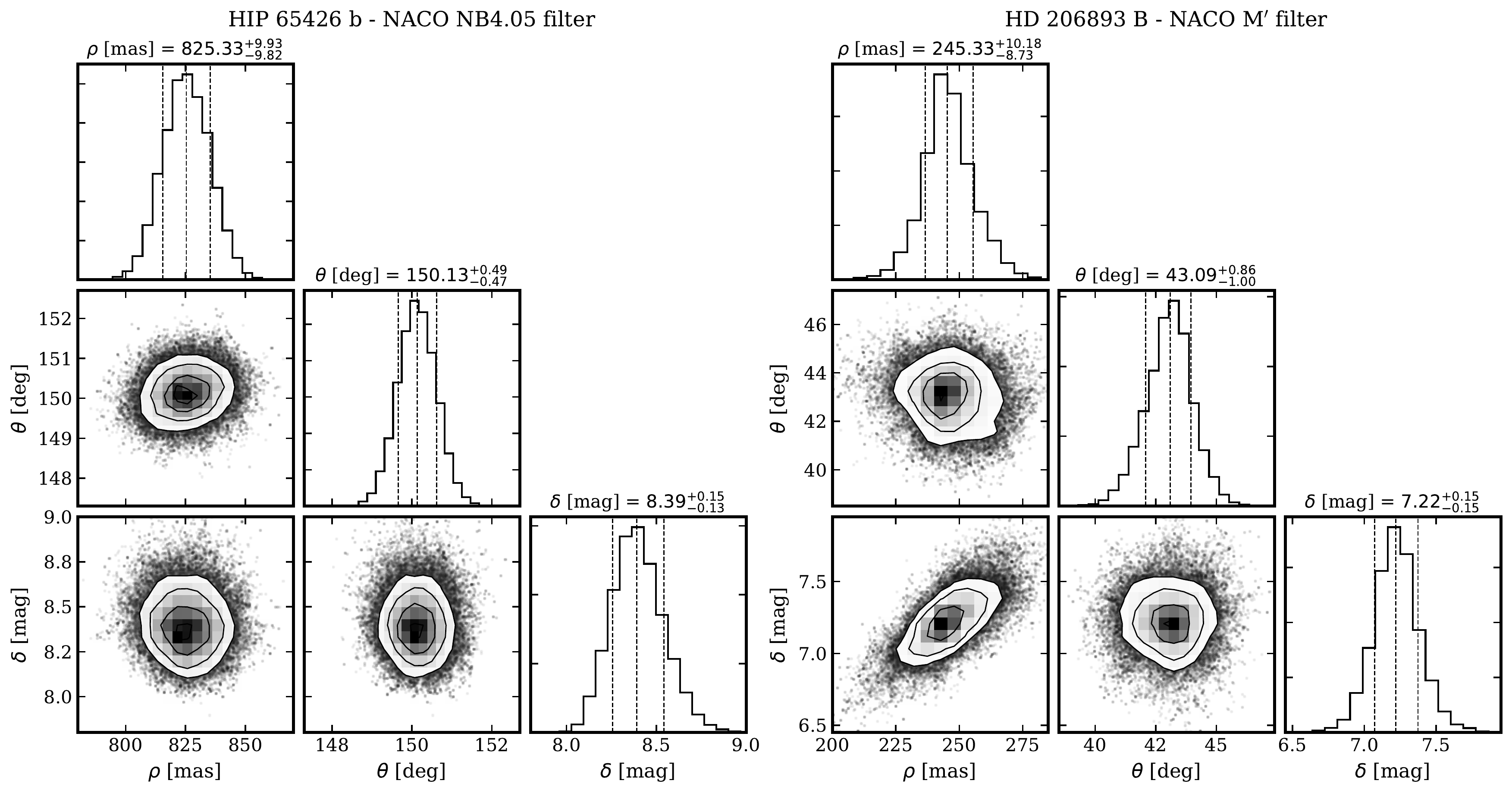}
\caption{Posterior distributions of the separation, $\rho$, position angle, $\theta$, and flux contrast, $\delta$, for HIP~65426~b in the NB4.05 filter (\emph{left panel}) and HD~206893 in the $M'$ filter (\emph{right panel}). The MCMC results of the remaining targets and filters are displayed in Fig.~\ref{fig:mcmc2} of Appendix~\ref{sec:appendix_posteriors}. The \emph{diagonal} panels show the marginalized 1D distributions of the parameters and the \emph{off-axis} panels map the 2D probability space for all parameter pairs. The listed values and uncertainties are the median, and the 16th and 84th percentiles of the parameter samples, which are also indicated by the vertically dashed lines in the 1D distributions. Contours overlaid on the 2D distributions correspond to $1\sigma$, $2\sigma$, and $3\sigma$ confidence levels for Gaussian statistics.\label{fig:mcmc1}}
\end{figure*}

We chose uniform priors for the three parameters and used the affine-invariance sampler implementation of \texttt{emcee} \citep{foreman2013} to compute the posterior distributions of the contrast and position of the companions. For each dataset, we let 200 walkers explore the probability landscape with 500 steps per walker (i.e., requiring $10^5$ PSF subtractions). The first 100 steps were removed after visual inspection of the walker's evolution. We then plotted the posterior distributions and adopted the median as the best-fit value and the 16th and 84th percentiles as the uncertainties. The mean acceptance fraction was typically $\sim$0.6--0.65 and the integrated autocorrelation time of each parameter in the range of $\sim$30--40 steps. Figure~\ref{fig:mcmc1} shows as an example the 1D and 2D marginalized distributions \citep[created with \texttt{corner.py};][]{foreman2016} of the NB4.05 dataset of HIP~65426~b and the $M'$ dataset of HD~206893~B. For the other datasets, we provide an overview of all remaining posterior distributions in Fig.~\ref{fig:mcmc2} of Appendix~\ref{sec:appendix_posteriors}.

\subsubsection{Error budget and measurement bias}\label{sec:error_budget}

The MCMC analysis provides an estimate of the statistical uncertainty of the companion contrast and position with the noise sampled from the distribution of reference pixels. In the derivation of the final contrast, we also include a correction for the intrinsic bias of the measurement and several systematic error components that are not captured by the likelihood function of the MCMC.

Azimuthal variations in the noise residuals from speckles and background flux may cause a bias in the retrieved flux contrast. The intrinsic offset and its uncertainty was estimated with the injection and retrieval of artificial planets. This was done by first removing the companion flux with the best-fit results from the MCMC analysis. We then injected a PSF template with the same contrast and separation as the real companion while the position angle was stepwise changed by $1\deg$ over the full $360\deg$. For each position, we retrieved the separation, position angle, and contrast of the artificial source by minimizing the $\chi^2$ function from Eq.~\ref{eq:likelihood}. The offset between injected and retrieved values were then calculated and we adopted the median as the bias of our measurement, and the 16th and 84th percentiles as the systematic uncertainty related to the remaining noise residuals after the PSF subtraction.

\begin{figure*}
\centering
\includegraphics[width=\linewidth]{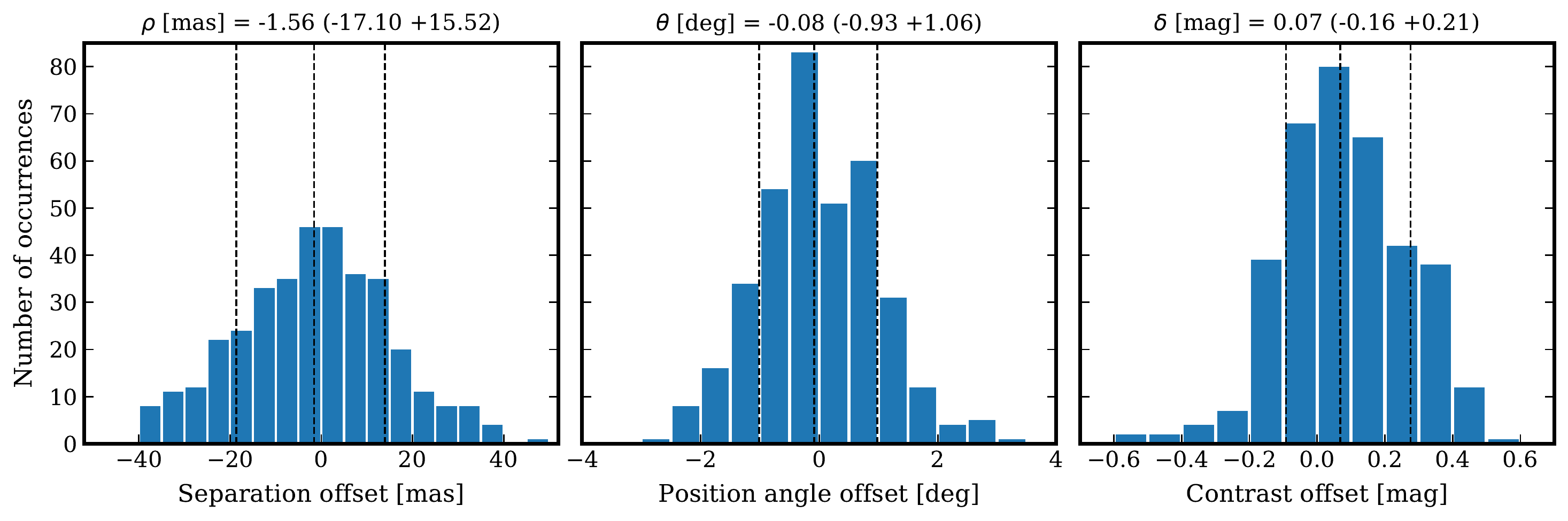}
\caption{Offset between the injected and retrieved values of the separation (\emph{left panel}), $\rho$, position angle (\emph{center panel}), $\theta$, and flux contrast (\emph{right panel}), $\delta$, for the NB4.05 dataset of HIP~65426. The precision of the figure of merit was tested at 360 position angles with the best-fit separation and contrast from the MCMC analysis (see Fig.~\ref{fig:mcmc1}). The vertically dashed lines indicate the 16th, 50th, and 84th percentiles of the samples from which the uncertainties in the title above each panel have been derived.\label{fig:error}}
\end{figure*}

Figure~\ref{fig:error} shows an example of the offset distributions for the NB4.05 dataset of HIP~65426. In this case, there is a bias in the contrast of 0.07~mag and the distribution of the offset is somewhat asymmetric with respect to the median. Specifically, the distribution is broader towards retrieved contrast values that are smaller than the injected value. Based on this analysis, we corrected the contrast by the bias offset and conservatively included the largest of the two error bars in the error budget (see Table~\ref{table:photometry}).

The bias of the contrast is typically small or even negligible except for the NB3.74 dataset of $\beta$~Pic and the $L'$ dataset of PZ~Tel. For these cases, we tested the use of different numbers of PCs, but this had only a small impact on the estimated bias and uncertainty. Interestingly, Fig.~\ref{fig:images} shows that in these two cases there is a prominent, negative residual present at the separation of the companion, likely related to an imperfect subtraction of the Airy pattern. Indeed, the $L'$ image in Fig.~\ref{fig:pztel} shows that the position of PZ~Tel~B coincides with a bright diffraction ring.

The stellar flux remained unsaturated throughout the observing sequences of the $M'$ data and also some of the narrowband data. In these cases, we applied a one-on-one injection for each image such that the calibration error related to the PSF template can be excluded from the error analysis. For the other datasets, we measured the stellar flux with a circular aperture (diameter of 2~FWHM) in each of the calibration exposures (see Sect.~\ref{sec:observations}). The standard deviation on these flux measurements was adopted as a measure for the variation in the stellar flux which degrades the precision with which the companion signal is removed in each individual image. For the $L'$ dataset of PZ~Tel, we also considered the uncertainty in the transmission of the neutral density filter that was used for the unsaturated exposures. We adopted a filter transmission of $(2.33 \pm 0.10)$\% from \citet{bonnefoy2013} and included the uncertainty in the error budget (see Table~\ref{table:photometry}).

The final contrast values and error components are listed in Table~\ref{table:photometry}. Since the MIRACLES program focuses on atmospheric characterization with 4--5~$\mu$m photometry, we do not analyze the astrometry of the companions. Nonetheless, the uncertainty on the separation and position angle were simultaneously computed with the uncertainty on the contrast. We have therefore listed the retrieved position values and error components in Table~\ref{table:astrometry} of Appendix~\ref{sec:appendix_astrometry}.

\begin{sidewaystable*}
\caption{Photometry and error budget.}
\label{table:photometry}
\centering
\bgroup
\def\arraystretch{1.25}
\begin{tabular}{L{2cm} L{1.2cm} C{1.6cm} C{1.8cm} C{0.9cm} C{1.3cm} C{1.7cm} C{1.7cm} C{1.9cm} C{2.2cm} C{3cm}}
\hline\hline
Target & Filter & MCMC contrast\tablefootmark{a} & Bias offset\tablefootmark{a} & Calib. error\tablefootmark{b} & ND filter error & Final contrast\tablefootmark{c} & Star magnitude & Apparent magnitude & Absolute magnitude & Flux \\
 & & (mag) & (mag) & (mag) & (mag) & (mag) & (mag) & (mag) & (mag) & (W m$^{-2}$ $\mu$m$^{-1}$) \\
\hline

$\beta$ Pic b & NB3.74 & $8.13 \pm 0.11$ & $-0.33 \pm 0.24$ & - & - & $7.80 \pm 0.26$ & $3.45 \pm 0.01$\tablefootmark{d} & $11.25 \pm 0.26$ & $9.77 \pm 0.26$ & $1.70 \pm 0.42$ $\times$ $10^{-15}$ \\
$\beta$ Pic b & NB4.05 & $7.54 \pm 0.02$ & $-0.01 \pm 0.04$ & - & - & $7.53 \pm 0.04$ & $3.45 \pm 0.01$\tablefootmark{d} & $10.98 \pm 0.04$ & $9.50 \pm 0.05$ & $1.61 \pm 0.07$ $\times$ $10^{-15}$ \\
HIP 65426 b & $L'$ & $8.56 \pm 0.06$ & $0.01 \pm 0.08$ & 0.07 & - & $8.56 \pm 0.12$ & $6.76 \pm 0.04$\tablefootmark{e} & $15.32 \pm 0.13$ & $10.13 \pm 0.13$ & $3.90 \pm 0.46$ $\times$ $10^{-17}$ \\
HIP 65426 b & NB4.05 & $8.39 \pm 0.15$ & $0.07 \pm 0.21$ & 0.03 & - & $8.46 \pm 0.26$ & $6.77 \pm 0.04$\tablefootmark{f} & $15.23 \pm 0.26$ & $10.04 \pm 0.26$ & $3.22 \pm 0.78$ $\times$ $10^{-17}$ \\
HIP 65426 b & $M'$ & $7.83 \pm 0.16$ & $0.06 \pm 0.24$ & - & - & $7.89 \pm 0.29$ & $6.76 \pm 0.04$\tablefootmark{g} & $14.65 \pm 0.29$ & $9.46 \pm 0.29$ & $2.99 \pm 0.82$ $\times$ $10^{-17}$ \\
PZ Tel B & $L'$ & $4.94 \pm 0.13$ & $-0.16 \pm 0.12$ & 0.11 & 0.03 & $4.78 \pm 0.21$ & $6.26 \pm 0.05$\tablefootmark{e} & $11.04 \pm 0.22$ & $7.68 \pm 0.22$ & $2.01 \pm 0.41$ $\times$ $10^{-15}$ \\
PZ Tel B & NB4.05 & $4.69 \pm 0.01$ & $-0.00 \pm 0.02$ & 0.04 & - & $4.69 \pm 0.05$ & $6.25 \pm 0.05$\tablefootmark{f} & $10.94 \pm 0.07$ & $7.57 \pm 0.07$ & $1.67 \pm 0.11$ $\times$ $10^{-15}$ \\
PZ Tel B & $M'$ & $4.63 \pm 0.01$ & $-0.00 \pm 0.02$ & - & - & $4.63 \pm 0.03$ & $6.30 \pm 0.02$\tablefootmark{g} & $10.93 \pm 0.03$ & $7.56 \pm 0.03$ & $9.18 \pm 0.28$ $\times$ $10^{-16}$ \\
HD 206893 B & $L'$ & $8.28 \pm 0.12$ & $-0.02 \pm 0.28$ & 0.02 & - & $8.27 \pm 0.30$ & $5.53 \pm 0.07$\tablefootmark{e} & $13.80 \pm 0.31$ & $10.74 \pm 0.31$ & $1.59 \pm 0.46$ $\times$ $10^{-16}$ \\
HD 206893 B & NB4.05 & $7.62 \pm 0.21$ & $0.01 \pm 0.32$ & 0.08 & - & $7.63 \pm 0.39$ & $5.53 \pm 0.07$\tablefootmark{f} & $13.16 \pm 0.40$ & $10.11 \pm 0.40$ & $2.16 \pm 0.81$ $\times$ $10^{-16}$ \\
HD 206893 B & $M'$ & $7.22 \pm 0.15$ & $0.01 \pm 0.27$ & - & - & $7.23 \pm 0.31$ & $5.54 \pm 0.07$\tablefootmark{g} & $12.77 \pm 0.32$ & $9.72 \pm 0.32$ & $1.68 \pm 0.50$ $\times$ $10^{-16}$ \\

\hline
\end{tabular}
\egroup
\tablefoot{\\
\tablefoottext{a}{For asymmetric uncertainties, we conservatively use the largest of the two values that were derived from the MCMC and bias analysis (see Figs.~\ref{fig:mcmc1}, \ref{fig:error}, and \ref{fig:mcmc2}).}\\
\tablefoottext{b}{The uncertainty related to the variable observing conditions which may bias the relative flux calibration in cases where the unsaturated PSF was only sampled at the start and/or end of the observations.}\\
\tablefoottext{c}{The final contrast is calculated by adding the bias offset and combining the error components in quadrature.}\\
\tablefoottext{d}{The $L'$ and $M'$ photometry ($3.454 \pm 0.003$~mag and $3.458 \pm 0.009$~mag, respectively) in the ESO photometric system have been adopted from \citet{bouchet1991}. The errors have been inflated to 0.01~mag to account for a minor color offset with the VLT/NACO filters.}\\
\tablefoottext{e}{The $L'$ magnitude has been adopted from the WISE $W1$ photometry (see top left panel of Fig.~\ref{fig:wise_naco}).}\\
\tablefoottext{f}{The NB4.05 magnitude has been adopted from the WISE $W1$ photometry with a minor color correction of $0.01$~mag for HIP~65426 and $-0.01$~mag for PZ~Tel (see top right panel of Fig.~\ref{fig:wise_naco}).}\\
\tablefoottext{g}{The $M'$ magnitude has been adopted from the WISE $W1$ magnitude with a minor color correction of $0.04$~mag for PZ~Tel and $0.01$~mag for HD~206893 (see bottom left panel of Fig.~\ref{fig:wise_naco}).}\\
}
\end{sidewaystable*}

\subsubsection{Absolute photometric calibration}\label{sec:absolute_calibration}

With the results from the contrast measurements at hand, we determined the fluxes of the companions  by considering the apparent magnitudes of their host star. However, apart from $\beta$~Pic (see Table~\ref{table:targets}), we were not able to find $L'$, NB4.05, or $M'$ photometry for any of the stars in our sample. Since the stars are all young and therefore expected to be variable at optical and NIR wavelengths, we derived the stellar magnitudes directly from the available WISE photometry without considering shorter wavelengths. The uncertainty from the adopted WISE photometry has been included in the error budget.

The stellar fluxes for the NACO filters were all derived from the WISE $W1$ filter ($\lambda_0 = 3.4$~$\mu$m), which is most comparable to the NACO $L'$ filter (see Fig.~\ref{fig:transmission}). Although the WISE $W2$ filter ($\lambda_0 = 4.6$~$\mu$m)  more closely resembles the NACO $M'$ filter than $W1$, it was shown by \citet{avenhaus2012} that for targets with an apparent magnitude brighter than $\sim$6~mag, the recovered (because the WISE detector saturates at 6.7~mag) $W2$ photometry may not be reliable. Given the differences in wavelength coverage between the WISE and NACO filters, we used synthetic spectra to quantify the colors between these two filter systems.

We first computed synthetic colors from the BT-NextGen atmospheric models \citep{allard2012} for effective temperatures in the range of 4000-10000~K (see Fig.~\ref{fig:wise_naco} in Appendix~\ref{sec:appendix_calibration}). Surface gravity and metallicity effects are negligible for the stellar temperatures of our sample. The results show that the color of the WISE $W1$ and NACO $L'$ is negligible for temperatures greater than $\sim$4500~K and so we adopted the WISE $W1$ magnitudes as $L'$ photometry for the stars in our sample (except for $\beta$~Pic). The color of the WISE $W1$ and NACO NB4.05 filters is $\lesssim$0.02~mag in the considered temperature range above $\sim$4500~K. Therefore, the $W1$ magnitudes are also used for the NB4.05 photometry but we applied a color correction of $+0.01$~mag and $-0.01$~mag for HIP~65426 and PZ~Tel, respectively, as determined from their effective temperatures.

Figure~\ref{fig:wise_naco} shows that the deviation of the $W1$~--~$W2$ colors of our sample from the model predictions is largest for the brightest of the targets (HD~206893), which may indeed point to a problem with the $W2$ photometry. We therefore also adopt the $W1$ photometry for the $M'$ calibration of the host stars and apply a color correction of $+0.04$~mag and $+0.01$~mag for PZ~Tel and HD~206893, respectively, while this is not required for HIP~65426 due to its earlier spectral type.

\section{Results}\label{sec:results}

\subsection{Companion fluxes and colors}\label{sec:fluxes_colors}

\begin{table*}
\caption{Companion colors at 3--5~$\mu$m.}
\label{table:colors}
\centering
\bgroup
\def\arraystretch{1.25}
\begin{tabular}{L{2.2cm} C{2.2cm} C{2.2cm} C{2.2cm}}
\hline\hline
Target & $L'$~--~NB4.05 & $L'$~--~$M'$ & NB4.05~--~$M'$ \\
 & (mag) & (mag) & (mag) \\
\hline
$\beta$~Pic~b &  $0.32 \pm 0.08$ &  $0.20 \pm 0.13$ & $-0.12 \pm 0.13$ \\
HIP~65426~b   &  $0.10 \pm 0.25$ &  $0.68 \pm 0.31$ &  $0.58 \pm 0.36$ \\
PZ~Tel~B      &  $0.10 \pm 0.23$ &  $0.11 \pm 0.22$ &  $0.01 \pm 0.08$ \\
HD~206893~B   &  $0.63 \pm 0.46$ &  $1.02 \pm 0.41$ &  $0.39 \pm 0.43$ \\
\hline
\end{tabular}
\egroup
\end{table*}

The extracted flux contrasts, calibrated magnitudes, and uncertainties in $L'$, NB4.05, and $M'$ are listed in Table~\ref{table:photometry} together with the new NB3.74 photometry of $\beta$~Pic~b. In addition, for the analysis of $\beta$~Pic~b, we adopted the $L'$ and $M'$ magnitudes from \citet{stolker2019} because we applied a similar procedure for the photometric extraction and uncertainty estimation in that study.

The \emph{Gaia} DR2 parallax of each target was used to calculate its distance (see Table~\ref{table:targets}), which we then used to convert the apparent magnitude into an absolute magnitude while taking into account the uncertainty on the parallax (and hence the distance) in the error propagation:
\begin{equation}
\sigma_{M}^2 = \sigma_{m}^2 + \sigma_{d}^2 \left( \frac{5}{d \log{10}}\right)^2,
\end{equation}
where $\sigma_{M}$ is the uncertainty on the absolute magnitude, $\sigma_{m}$ the uncertainty on the apparent magnitude, $\sigma_{d}$ the uncertainty on the distance, and $d$ the distance in parsecs. The absolute magnitudes are listed in Table~\ref{table:photometry}, revealing some diversity in the sample since the values provide a measure for the temperature of the atmospheres.

Apparent magnitudes were converted to physical fluxes by calculating the zero point of the filters with a flux-calibrated spectrum of Vega \citep{bohlin2007} and setting its magnitude to 0.03~mag for all filters. The zero point of each filter is computed as
\begin{equation}\label{eq:mean_flux}
\left\langle f_\lambda \right\rangle = \frac{\int f_\lambda(\lambda) R(\lambda) d\lambda}{\int R(\lambda) d\lambda},
\end{equation}
where $\left\langle f_\lambda \right\rangle$ is the mean flux for a given filter, $f_\lambda(\lambda)$ the wavelength-dependent flux, and $R(\lambda)$ the filter transmission. From this we obtained a zero point of $5.21 \times 10^{-11}$~W~m$^{-2}$~$\mu$m$^{-1}$ in NB3.74, $5.12 \times 10^{-11}$~W~m$^{-2}$~$\mu$m$^{-1}$ in $L'$, $3.86 \times 10^{-11}$~W~m$^{-2}$~$\mu$m$^{-1}$ in NB4.05, and $2.10 \times 10^{-11}$~W~m$^{-2}$~$\mu$m$^{-1}$ in $M'$. We neglected additional components in the system response function such as telluric transmission, mirror reflectivity, optics transmission, and quantum efficiency, thereby assuming that these have a similar impact on the zero point and the companion's photometry, even though there will be dissimilarities between the spectral morphology of Vega and a substellar atmosphere. The calibrated fluxes are listed in Table~\ref{table:photometry}.

From the derived magnitudes, we calculated the colors of the companions. These are listed in Table~\ref{table:colors} with the errors calculated by adding the uncertainties of the individual filters in quadrature. The overview shows that many of the colors are within their uncertainties consistent with being blue, gray, or red. One exception is the $L'$~--~NB4.05 color of $\beta$~Pic~b, which is red at a $4\sigma$ confidence level. Furthermore, the red $L'$~--~NB4.05 color of HD~206893~B, the red NB4.05~--~$M'$ color of HIP~65426~b, and the red $L'$~--~$M'$ colors of $\beta$~Pic~b, HIP~65426~b, and HD~206893~B are all 1--2.5$\sigma$ results. A more detailed comparison and analysis of the magnitudes and colors will be provided below.

\subsection{Color and magnitude comparison: evolution, chemical composition, and clouds}\label{sec:color_magnitude}

The evolution of substellar-mass objects is marked by gravitational contraction which is only counteracted by electron degeneracy pressure. Consequently, brown dwarfs and gas-giant planets cool over time such that their luminosity has a strong dependence on age \citep[e.g.,][]{burrows2001}. The spectral sequence of substellar objects extends from M-type stars to L, T, and Y dwarfs, which at a given age corresponds to an isochrone of decreasing mass and temperature. This implies that for a given spectral type, substellar objects can have a different surface gravity while having a similar temperature as a result of having a different age. Therefore, we investigate the $L'$, NB4.05, and $M'$ color and magnitude characteristics of our sample of young directly imaged low-mass companions by comparing them with synthetic photometry from atmospheric models, photometry from field and young, low-gravity brown dwarfs, and other directly imaged companions.

\subsubsection{Selected empirical data and model spectra}\label{sec:magnitude_data}

The color and magnitude comparison was done with \texttt{species}, which is a toolkit that we developed for analyzing spectral and photometric data of planetary and substellar atmospheres. The software provides a coherent, easy-to-use framework to store, inspect, analyze, and plot observational data and models. It benefits from a wide variety of publicly available data such as atmospheric model spectra, photometric libraries, spectral libraries, evolutionary tracks, photometry of directly imaged companions, and filter transmission profiles. The Python package is publicly available in the PyPI repository\footnote{\url{https://pypi.org/project/species}} and maintained on Github\footnote{\url{https://github.com/tomasstolker/species}}. The online documentation\footnote{\url{https://species.readthedocs.io}} contains additional information on the workflow, implemented features, supported data, and several tutorials. Before presenting the results, we provide a brief summary of the empirical data and the model spectra that were selected.

The colors and magnitudes of the observed sample are empirically compared with the photometry from the Database of Ultracool Parallaxes \citep{dupuy2012,dupuy2013,liu2016}. This is an inventory of late-M, L, T, and Y dwarfs with measured parallaxes and photometry taken in the MKO filter system (including $L'$ and $M'$ magnitudes for a subset of the data). We extracted the field objects, which are old and are expected to have a high surface gravity, and also objects that were flagged as being young and/or having a low surface gravity.

While the NACO $L'$ and $M'$ filters are comparable to respective filters from the MKO system, the NB4.05 filter transmission deviates significantly from any of the available broadband photometry. We therefore used the IRTF spectral library \citep{cushing2005}, which contains a selection of M and early L dwarf spectra in the 3--4$\mu$m range. Specifically, these spectra extend up to 4.1~$\mu$m, where the NB4.05 and $L'$ filter transmission has dropped to 10\% and 65\%, respectively. Synthetic fluxes for the $L'$ and NB4.05 filters were therefore computed from these spectra while ignoring the slight deficit in the spectral range.

Spectra in the $L'$ and $M'$ band regime of late L and T dwarfs are more sparsely available due to their low luminosity and the high thermal background emission. We adopted the synthetic $L'$ and NB4.05 photometry from \citet{currie2014} which were calculated from spectra of brown dwarfs obtained with the AKARI infrared space telescope. In this sample, we included eleven L dwarfs and two early T dwarfs.

The photometry is also compared with predictions from evolutionary and atmospheric models. For this, we used the isochrone data from \citet{baraffe2003}, who coupled interior models to the non-gray atmospheric models by \citet{allard2001}. The AMES-Cond and AMES-Dusty models present two limiting cases for the treatment of the dust. The first model includes efficient gravitational settling of the dust with no effect on the thermal structure and the emitted spectrum. The second model considers inefficient gravitational settling with the grains remaining at the altitude where they form, following chemical equilibrium conditions. Isochrones were extracted at ages of 20~Myr, 100~Myr, and 1~Gyr and were interpolated for logarithmically spaced masses. From this, we obtained the effective temperature and surface gravity for each mass--age pair, and calculated the corresponding radius. We then used the spectra of the AMES-Cond and AMES-Dusty models to compute the synthetic photometry for the NACO filters.

Finally, we also gathered photometry of any directly imaged companions with magnitudes available in at least two of the $H$, $L'$, NB4.05, and $M'$ filters. The comparison includes photometry from $\beta$~Pic~b \citep{currie2013}, HIP~65426~b \citep{chauvin2017}, PZ~Tel~B \citep{biller2010}, HD~206893~B \citep{milli2017}, HR~8799~bcde \citep{galicher2011,currie2012,zurlo2016,currie2014}, GSC~06214~B \citep{ireland2011,bailey2013}, ROXs~42~Bb \citep{daemgen2017}, 51~Eri~b \citep{rajan2017}, kappa~And~b \citep{bonnefoy2014}, 2M1207~b \citep{chauvin2004}, 2M0103~ABb \citep{delorme2013}, and 1RXS~1609~B \citep{lafreniere2008,lafreniere2010}. Here, we loosely combined broadband filters from different instruments, as well as several magnitudes from the narrowband VLT/SPHERE $H2$ filter. Small color offsets up to a few tenths of a magnitude are therefore to be expected but this will have no major impact on the overall picture emerging from these diagrams.

\subsubsection{Color--magnitude diagrams}\label{sec:color_mag_diagrams}

\begin{figure*}
\centering
\includegraphics[width=\linewidth]{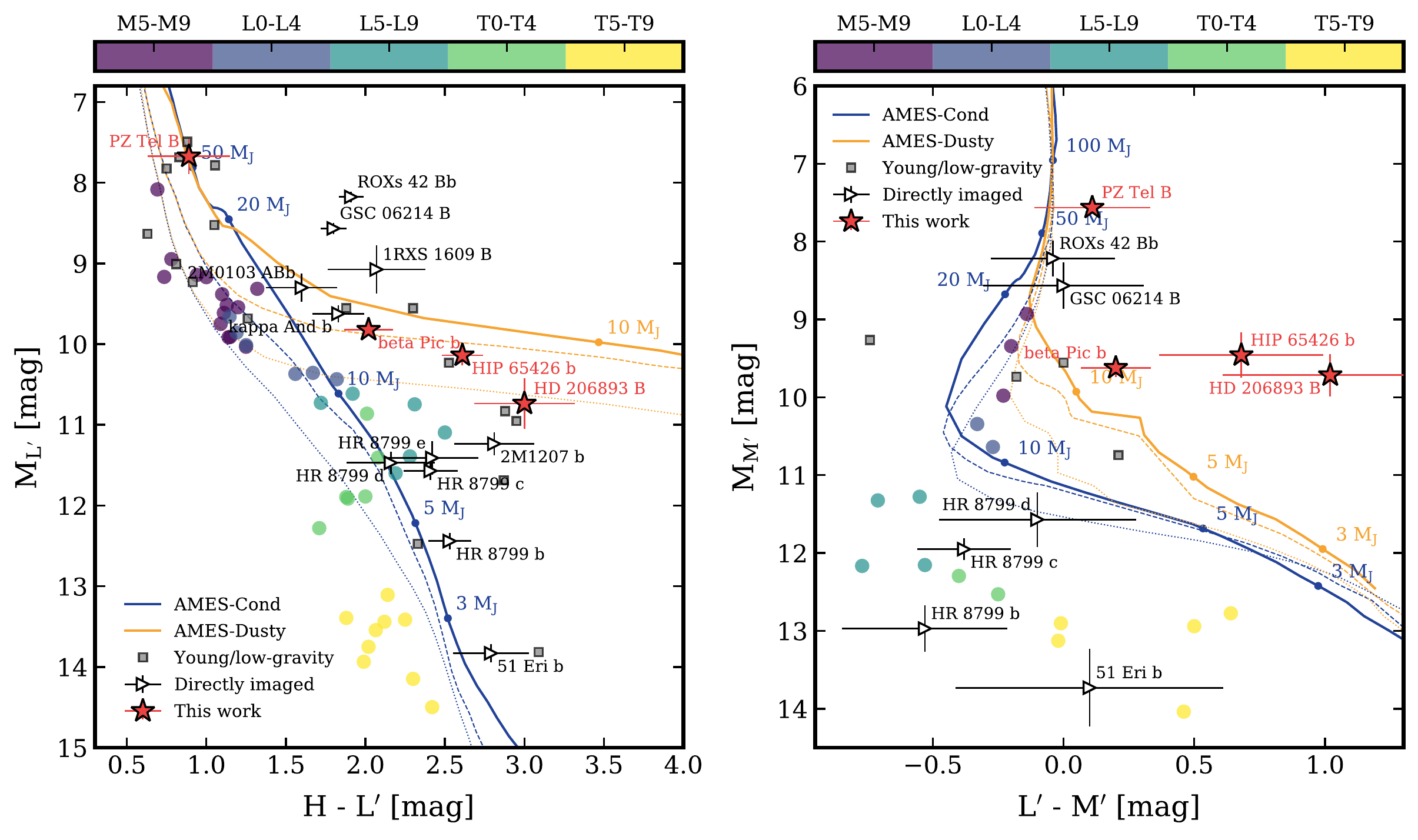}
\caption{Color--magnitude diagrams of $H$~--~$L'$ vs. M$_{L'}$ (\emph{left panel}) and $L'$~--~$M'$ vs. M$_{M'}$ (\emph{right panel}). The field objects are color coded by M, L, and T spectral types (see discrete color bar), the young and/or low-gravity dwarf objects are indicated with a \emph{gray square}, and the directly imaged companions are labeled individually. The companions from this study are highlighted with a \emph{red star}. The \emph{blue} and \emph{orange} lines show the synthetic colors and magnitudes computed from the AMES-Cond and AMES-Dusty evolutionary tracks for ages of 20~Myr (\emph{solid}), 100~Myr (\emph{dashed}), and 1000~Myr (\emph{dotted}).\label{fig:color_mag1}}
\end{figure*}

Color--magnitude diagrams illustrate the luminosity evolution of substellar atmospheres through the correlation between spectral type and absolute flux. Figure~\ref{fig:color_mag1} presents two such diagrams based on the $H$, $L'$, and $M'$ photometry of the companions in comparison with field and young/low-gravity dwarfs and a sample of directly imaged companions.

The absolute $L'$ magnitudes and $H$~--~$L'$ colors of the directly imaged companions appear to follow a similar track as the young/low-gravity objects (see left panel in Fig.~\ref{fig:color_mag1}). While the field and young/low-gravity dwarfs share the same characteristics at the high-mass end of the diagram, the young/low-gravity objects are redder compared to their older counterparts within the L and T-type regime of the sequence. A similar trend can be identified in the photometric characteristics of the studied sample. PZ~Tel~B coincides with the photometry and colors from the isolated brown dwarfs while $\beta$~Pic~b, HIP~65426~b, and HD~206893~B are red in $H$~--~$L'$ compared to the field dwarfs but follow the line of expectation with respect to the the young/low-gravity objects. In contrast to the colors, the absolute $L'$ fluxes of the sample follow approximately those of the field dwarfs if we consider the following spectral types: M7$\pm$1 for PZ~Tel~B \citep{maire2016}, L2$\pm$1 for $\beta$~Pic~b \citep{chilcote2017}, L5--L7 for HIP~65426~b \citep{chauvin2017}, and late~L for HD~206893~B \citep{delorme2017}.

The atmospheric models predict comparable values at the high end of the spectral sequence where the temperatures are too high for cloud species to condense. Towards lower mass planets, the $H$~--~$L'$ color becomes increasingly red due to the broad H$_2$O absorption features at 1.4~$\mu$m and 1.8~$\mu$m that partially covers the $H$ band regime. The CH$_4$ absorption at 3.3~$\mu$m in the AMES-Cond (clear atmosphere) spectra starts to affect the $L'$ photometry in the warm temperature regime. The combined effect causes the $H$~--~$L'$ color to remain constant while the absolute $L'$ flux decreases until the absorption by H$_2$O takes the overhand for the lowest planet masses. The AMES-Dusty (cloudy atmosphere) spectra show a weak CH$_4$ feature around 3.3~$\mu$m but the color in the warm and cold temperature regime is mostly affected by the strongly mixed dust grains. This continuum source shifts the overall flux towards longer wavelengths, hence the increasingly red color.

The right panel of Fig.~\ref{fig:color_mag1} shows the $L'$~--~$M'$ versus M$_{M'}$ color--magnitude diagram. In this case, there is a smaller sample of directly imaged companions and isolated objects available. The colors of $\beta$~Pic~b, HIP~65426~b, and HD~206893~B are redder than both the field and young/low-gravity dwarfs (although with marginal significance), as well as the synthetic colors from the atmospheric models. Interestingly, the absolute $M'$ magnitudes of $\beta$~Pic~b, HIP~65426~b, and HD~206893~B are all comparable ($\sim$9.5~mag) and similar to the magnitudes of the late M type field objects. That is, these three directly imaged objects are overluminous in $M'$ given their spectral types. The color of PZ~Tel~B on the other hand is within its uncertainty consistent with the model predictions, but its absolute $M'$ flux is potentially also slightly enhanced.

Absorption by CO in the $M'$ band causes the color from the AMES-Cond models to become blue for the high-temperature objects, after which the CH$_4$ absorption in the $L'$ band increases in its relative strength, resulting in a red color. The models overpredict the $M'$ flux of the late L and T dwarf sequence because the abundances of CO and CH$_4$ in their atmospheres are likely driven by nonequilibrium chemistry due to vertical mixing of CO from the deeper convective layers \citep[e.g.,][]{griffith1999,yamamura2010}. Enhanced CO absorption in the $M'$ band will lead to a bluer color and a lower flux in $M'$. The AMES-Dusty models predict colors that are mostly gray at high temperatures and increasingly red towards lower temperatures due the condensation of dust grains. The $M'$ flux and color predicted by these models reaches closer to the observed values from $\beta$~Pic~b, HIP~65426~b, and HD~206893 but a significant discrepancy remains.

\begin{figure}
\centering
\includegraphics[width=\linewidth]{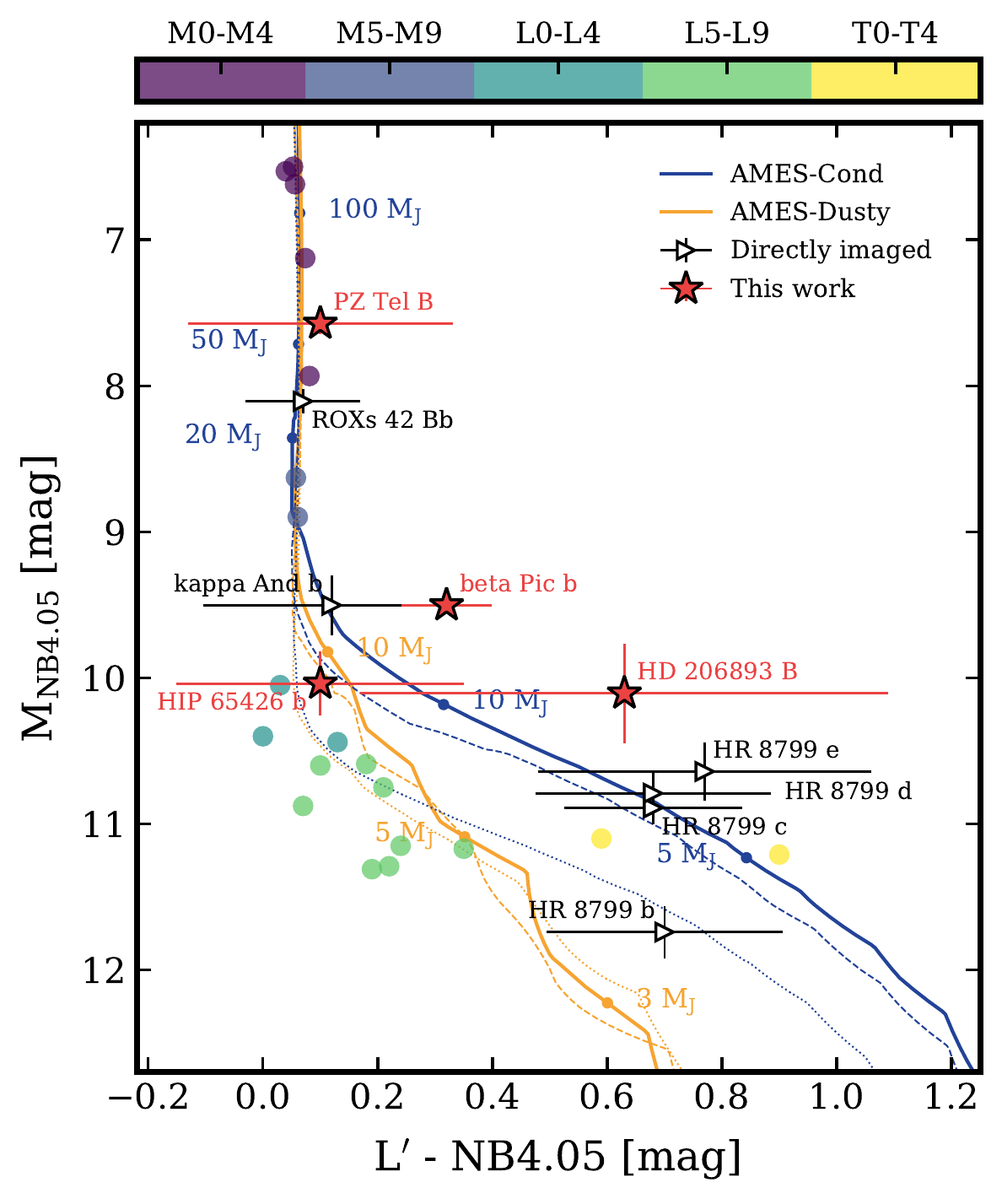}
\caption{Color--magnitude diagram $L'$~--~NB4.05 vs. M$_\textrm{NB4.05}$. The photometry and colors of the M, L, and T dwarfs were derived from IRTF \citep{cushing2005} and AKARI spectra \citep{currie2014}. The field objects are color coded according to their spectral type (see discrete color bar) and the directly imaged companions are labeled individually. The companions from this study are highlighted with a \emph{red star}. The \emph{blue} and \emph{orange} lines show the synthetic colors and magnitudes computed from the AMES-Cond and AMES-Dusty evolutionary tracks for ages of 20~Myr (\emph{solid}), 100~Myr (\emph{dashed}), and 1000~Myr (\emph{dotted}).\label{fig:color_mag2}}
\end{figure}

The $L'$~--~NB4.05 color in Fig.~\ref{fig:color_mag2} varies moderately with values ranging up to only 1~mag for the directly imaged objects close to the L/T transition. PZ~Tel~B and HIP~65426~b have a color that is consistent with the field dwarfs and model predictions. HD~206893~B and $\beta$~Pic~b on the other hand are redder across these wavelengths, although the uncertainty on the color of HD~206893~B is large. Its color is comparable with the HR~8799 planets but these objects are intrinsically fainter at 4~$\mu$m. Similar to the $M'$ magnitudes, all objects in our sample are brighter in NB4.05 compared to field dwarfs with similar spectral types.

The NB4.05 filter covers a regime with limited molecular absorption while CH$_4$ absorption at 3.3~$\mu$m can be present in the blue end of the $L'$ filter. As a result, the AMES-Cond spectra show a color that changes from gray to red. In the AMES-Dusty spectra, the CH$_4$ absorption feature is much weaker and it is mostly the dust opacity that causes a reddening of the spectra, albeit weaker than the reddening seen in the $H$~--~$L'$ colors.

\subsubsection{Color--color diagrams}\label{sec:color_color_diagrams}

\begin{figure*}
\centering
\includegraphics[width=\linewidth]{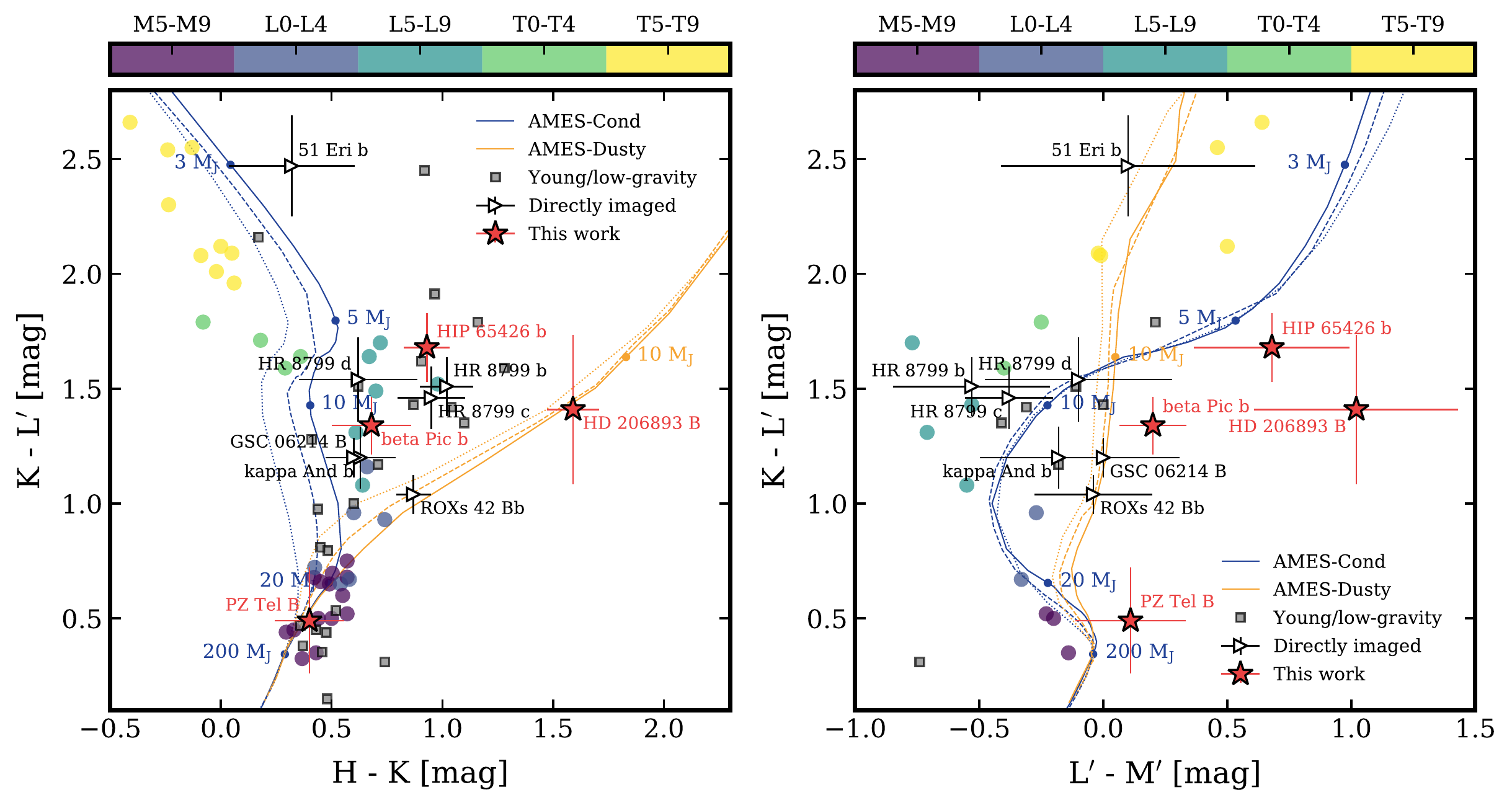}
\caption{Color--color diagrams for $H$~--~$K$ vs. $K$~--~$L'$ (\emph{left panel}) and $L'$~--~$M'$ vs. $K$~--~$L'$ (\emph{right panel}). The field objects are color coded into M, L, and T spectral types (see discrete color bar), the young and/or low-gravity dwarf objects are indicated with a \emph{gray square}, and the directly imaged companions are labeled individually. The companion colors derived in this study are highlighted with a \emph{red star}. The \emph{blue} and \emph{orange} lines are the synthetic colors that have been calculated from the AMES-Cond and AMES-Dusty evolutionary tracks for ages of 20~Myr (\emph{solid}), 100~Myr (\emph{dashed}), and 1000~Myr (\emph{dotted}).\label{fig:color_color}}
\end{figure*}

Color--color diagrams reveal correlations between the spectral slopes of two wavelength regimes. The left panel of Fig.~\ref{fig:color_color} shows the relation between the $H$~--~$K$ and $K$~--~$L'$ colors. At high temperatures, the field and young/low-gravity dwarfs show a similar correlation, which is also followed by PZ~Tel~B and $\beta$~Pic~b. For late L spectral types on the other hand, the colors of the young/low-gravity objects start to deviate from the field dwarfs. Regarding our sample, HIP~65426~b lies in the color regime of the young/low-gravity objects while HD~206893~B has more distinctive color characteristics.

The synthetic colors computed from the AMES-Cond and AMES-Dusty models show a divergence at masses below 20~$M_\mathrm{Jup}$. A clear atmosphere causes a blue color towards lower temperatures because of the decreasing $K$ band flux while at higher temperatures the $H$~--~$K$ color remains approximately constant because of the H$_2$O absorption at 1.4~$\mu$m, 1.8~$\mu$m, and 2.6~$\mu$m. The $K$~--~$L'$ color is close to gray at high temperatures and becomes redder at lower temperatures due to strong H$_2$O absorption in the $K$ band even though the CH$_4$ absorption also increases in the $L'$ band. For the AMES-Dusty model, the absorption features are largely muted and the increasingly red $H$~--~$K$ color is mainly caused by the dust continuum opacity.

The color--color relation between $K$~--~$L'$ and $L'$~--~$M'$ shows a larger dispersion when comparing the field and young/low-gravity dwarfs with the directly imaged companions. While the $K$~--~$L'$ colors are comparable for these two samples (see left panel of Fig.~\ref{fig:color_color}), the $L'$~--~$M'$ colors of the young/low-gravity objects appear red compared to the field dwarfs. Interestingly, the four directly imaged companions from our sample are even redder compared to the isolated objects except for PZ~Tel~B. The photometric characteristics of PZ~Tel~B are within their uncertainties comparable to those of field dwarfs and model predictions. The offset between the model colors and the field dwarfs is caused by an under- and overprediction of the $L'$ and $M'$ flux, respectively. This is related to the nonequilibrium CH$_4$ and CO abundances, and similar to the discrepancy in the color--magnitude diagram of $L'$~--~$M'$ versus M$_{M'}$ (see Fig.~\ref{fig:color_mag1}).

\subsection{Modeling of $\beta$~Pic~b with a dynamical mass prior}\label{sec:beta_pic}

The surface gravity of an atmosphere is an important parameter because it influences the vertical distribution of both the gas and the cloud condensates. A lower surface gravity will shift the photosphere to lower pressures and temperatures. Consequently, pressure broadening of the absorption lines becomes weaker and the atmosphere more transparent. However, quantifying the surface gravity from photometry alone is challenging and typically requires high-precision spectra.

Several studies of $\beta$~Pic~b have carried out a detailed atmospheric characterization to constrain the planet's main physical properties \citep[e.g.,][]{morzinski2015,chilcote2017}. Since the dynamical mass of $\beta$~Pic~b was recently determined \citep{snellen2018,dupuy2019}, we carried out a simplified fitting procedure, only considering photometry of the planet but also applying a prior on its mass. The analysis was done with \texttt{species} with which we used a grid of synthetic spectra from the DRIFT-PHOENIX model \citep{helling2008}. This is a radiative-convective equilibrium atmosphere model which includes a detailed cloud model for predicting the composition and size distribution of the condensates.

\begin{figure*}
\centering
\includegraphics[width=0.9\linewidth]{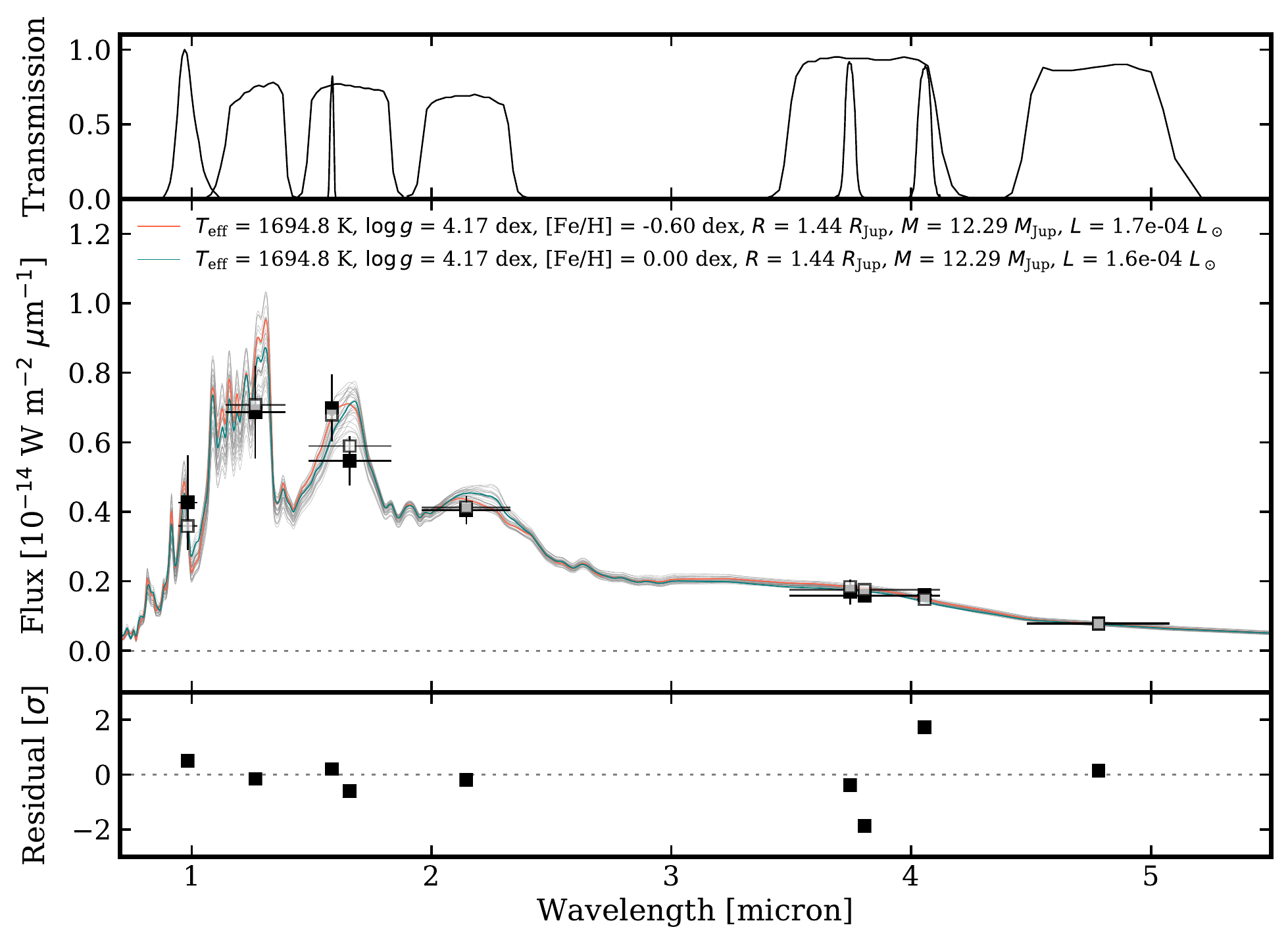}
\caption{Synthetic spectra ($\lambda/\Delta\lambda = 50$) of the DRIFT-PHOENIX atmospheric models that best describe the photometry of $\beta$~Pic~b with a prior on its mass. The two colored lines show the interpolated spectra for the median values of the posterior distributions, except for the metallicity, which is set to $\mathrm{[Fe/H] = -0.06}$ (\emph{red}) and $\mathrm{[Fe/H] = 0.0}$ (\emph{teal}). Additionally, 30 sets of parameter values were randomly drawn from the posterior distributions (see Fig.~\ref{fig:betapic_posterior}) and shown as \emph{gray} spectra. The \emph{black solid squares} are the data and the \emph{black open squares} are the synthetic photometry for the best-fit model. The differences between the data and the synthetic photometry are shown in multiples of the 1$\sigma$ uncertainties. The horizontal error bars indicate the FWHM of the filter transmission profiles, which are also plotted above the spectra. The mass and luminosity in the legend have been computed from the fit results.\label{fig:betapic_spectrum}}
\end{figure*}

The posterior distributions of the effective temperature, $T_\mathrm{eff}$, surface gravity, $\log{g}$, metallicity, $\mathrm{[Fe/H]}$, and radius, $R$, were computed with the affine-invariant sampler \texttt{emcee} \citep{foreman2013}. For each sample, the grid of spectra was linearly interpolated in the multidimensional space and synthetic photometry was computed with Eq.~\ref{eq:mean_flux}. The log-likelihood function was then calculated as
\begin{equation}
\log \mathcal{L} \propto -\frac{1}{2} \sum_\mathrm{i}^N \left(\frac{f_\mathrm{i}-m_\mathrm{i}}{\sigma_\mathrm{i}}\right)^2,
\end{equation}
where $f_\mathrm{i}$ is the flux in filter $i$, $m_\mathrm{i}$ is the synthetic flux computed from the model spectra in filter $i$, and $\sigma_\mathrm{i}$ is the uncertainty on the observed flux. We used 200 walkers, each making a 1000 steps, which were initialized at random positions close to the approximate temperature and radius, while the surface gravity and metallicity were uniformly initialized at random values between the grid boundaries.

The prior on the surface gravity was calculated by assuming a Gaussian prior distribution for the mass, which was centered at the model-independent value of $13 \pm 3$~$M_\mathrm{Jup}$ from \citet{dupuy2019}. A uniform prior was chosen for all other parameters. Although $\beta$~Pic~b has been routinely observed by various instruments, we only considered the following photometry without duplicating filter bandpasses: Magellan/VisAO $Y_\mathrm{s}$ filter \citep{males2014}, VLT/NACO $J$ filter \citep{currie2013}, Gemini/NICI $CH_\mathrm{4S,1\%}$ \citep{males2014}, VLT/NACO $H$ filter \citep{currie2013}, VLT/NACO $K_\mathrm{s}$ filter \citep{bonnefoy2011}, VLT/NACO $L'$ and $M'$ \citep{stolker2019}, and VLT/NACO with NB3.74 and NB4.05 filters from this work.

Half of the walker's steps were discarded as burn-in, leaving a total of $10^5$ samples. From these samples, we computed 30 randomly selected spectra which are shown in Fig.~\ref{fig:betapic_spectrum} together with the best-fit spectra for the two different metallicity values (see below). The residuals between the observed fluxes and the synthetic photometry are all within $2\sigma$ from the expected values. A dispersion of the spectra is mostly visible at the shorter wavelengths given the uncertainties on the photometry in the 1--2~$\mu$m region.

\begin{figure*}
\centering
\includegraphics[width=0.9\linewidth]{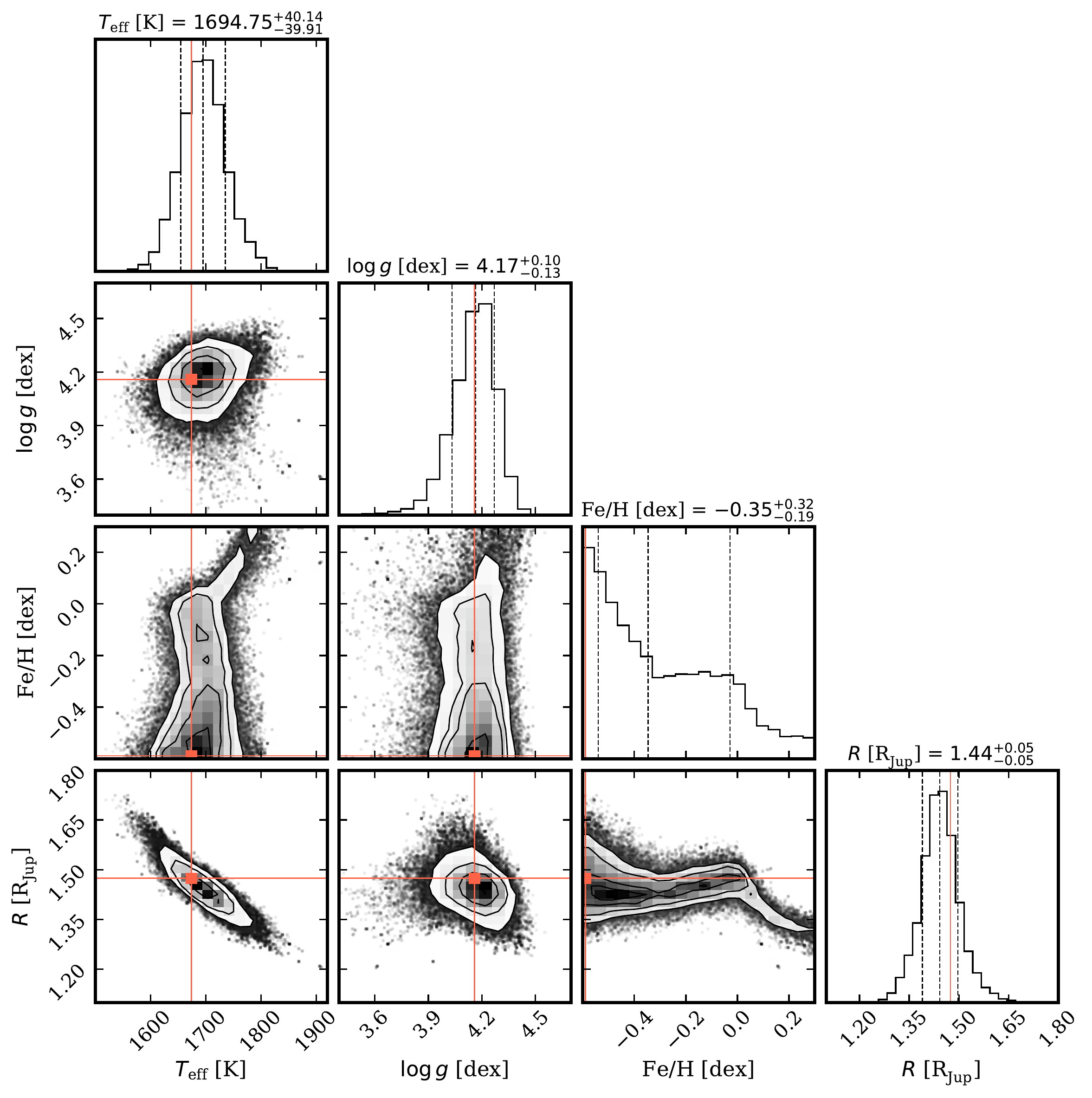}
\caption{Posterior distributions of the atmospheric model parameters that were fitted to the photometry of $\beta$~Pic~b with a prior on its mass. A comparison of the photometry and the sampled model spectra is shown in Fig.~\ref{fig:betapic_spectrum}. The marginalized 1D and 2D distributions are shown together with the median, and 16th and 84th percentiles of the samples. The \emph{red} horizontal and vertical lines indicate the sample position with the maximum posterior probability.\label{fig:betapic_posterior}}
\end{figure*}

The sampled posterior distributions are shown in Fig.~\ref{fig:betapic_posterior} \citep{foreman2016} with the median value of the samples indicated above each of the 1D distributions. Without the prior on the mass of the planet, the surface gravity was constraint to only $\log{g} = 4.29_{-0.76}^{+0.67}$~dex while showing a strong degeneracy with the metallicity of the atmosphere. With the prior on the mass, we constrained the surface gravity to a value of $\log{g} = 4.17_{-0.13}^{+0.10}$~dex. The metallicity remained uncertain since its effect on the spectral appearance is relatively small compared to the uncertainties on the photometry. The posterior distribution of the metallicity has a peak both at -0.6~dex (at the lower edge of the grid) and a smaller peak close to solar metallicity. The effective temperature and radius are not significantly affected by the mass prior and were already constrained with the photometry alone. The best-fit values derived from the posterior distributions are $T_\mathrm{eff} = 1694 \pm 40$~K and $R = 1.44 \pm 0.05$~$R_\mathrm{Jup}$.

\subsection{Detection limits in NB4.05 and $M'$ filters}\label{sec:limits}

The detection limits of our data were calculated using a method that is similar to the PSF subtraction described in Sect.~\ref{sec:data_processing}. First, the azimuthal noise level was computed for a given position that was tested. An artificial planet was then injected with a S/N of 100 after which the self-subtraction inherent to PCA was determined by measuring the signal of the artificial source with a 1~FWHM aperture before and after the PSF subtraction. The contrast was then scaled to a false positive fraction (FPF) of $2.86 \times 10^{-7}$, which corresponds to a $5\sigma$ detection in the limit of Gaussian statistics. Since the noise measurement is limited by the small number of samples (i.e., the number of nonoverlapping apertures at a given separation), we assume a Student's \emph{t}-distribution for the calculation of the FPF \citep{mawet2014}. The positions at which the detection limits were calculated started at an inner separation of 200~mas and increased in steps of 10~mas while the position angle was uniformly sampled with six positions. The average of the six azimuthal positions was stored and the procedure was repeated for a range of 5--30~PCs in steps of 5~PCs.

\begin{figure*}
\centering
\includegraphics[width=\linewidth]{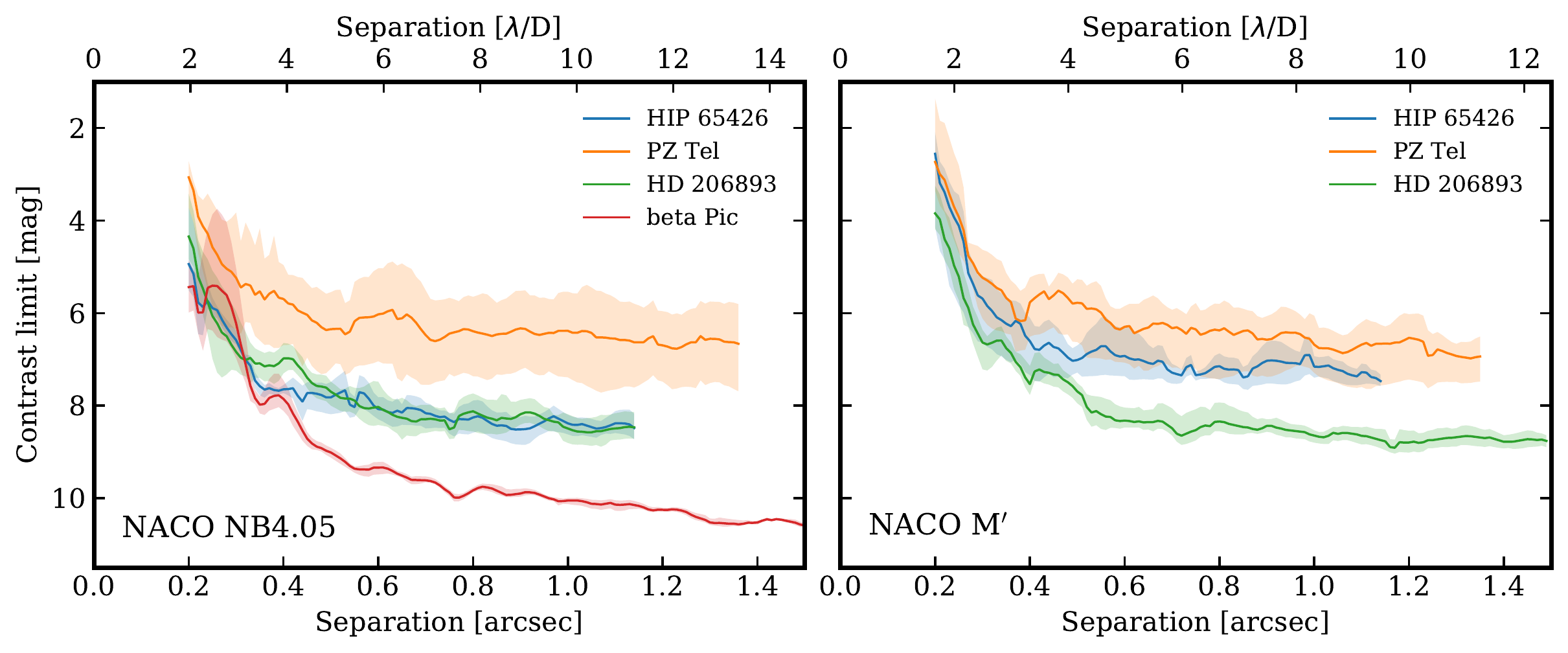}
\caption{Detection limits derived for the observed targets in the NACO NB4.05 (\emph{left panel}) and $M'$ (\emph{right panel}) filters. The limits for the archival dataset of $\beta$~Pic~b in NB4.05 are also included as a reference. The \emph{solid lines} show the mean contrast from six values of principal components (PCs) that were tested (5, 10, 15, 20, 25, and 30). The \emph{shaded area} covers the range between the minimum and maximum detection limit from the different number of PCs. The upper horizontal axis is shown in units of $\lambda/D$, that is, 102~mas in NB4.05 and 120~mas in $M'$ (at the central wavelengths of the filters).\label{fig:limits}}
\end{figure*}

The results are presented in Fig.~\ref{fig:limits} for the NB4.05 and $M'$ filters. The dispersion due to the different numbers of PCs that were used is largest at small separations where the impact of self-subtraction effects varies more strongly as function of PCs. The detection limits follow a similar steepness in the speckle-limited regime where field rotation causes only a small amount of movement of a signal while the curves quickly flatten off to the background-limited regime. This regime appears to start approximately at $\sim$8$\lambda/D$ for the NB4.05 data (except for $\beta$~Pic due to its brightness) and already at $\sim$5$\lambda/D$ for the $M'$ data. The limiting apparent magnitude in NB4.05 that was reached in the background-limited regime is approximately $12.5$~mag for PZ~Tel (16.8~min), $15.2$~mag for HIP~65426 (2.6~hr), $14.3$~mag for HD~206893 (48.9~min), and $14.3$~mag for $\beta$~Pic (29.7~min). For the $M'$ data, the limiting magnitude is about $13.3$~mag for PZ~Tel (24.4~min), $14.2$~mag for HIP~65426 (54.9~min), and $14.3$~mag for HD~206893 (73.9~min), where values between parentheses indicate the effective integration time after application of the frame selection.

\section{Discussion and future outlook}\label{sec:discussion}

\subsection{Atmospheric insights from 3--5~$\mu$m photometry}\label{sec:atmospheric_insights}

\begin{figure*}
\centering
\includegraphics[width=0.8\linewidth]{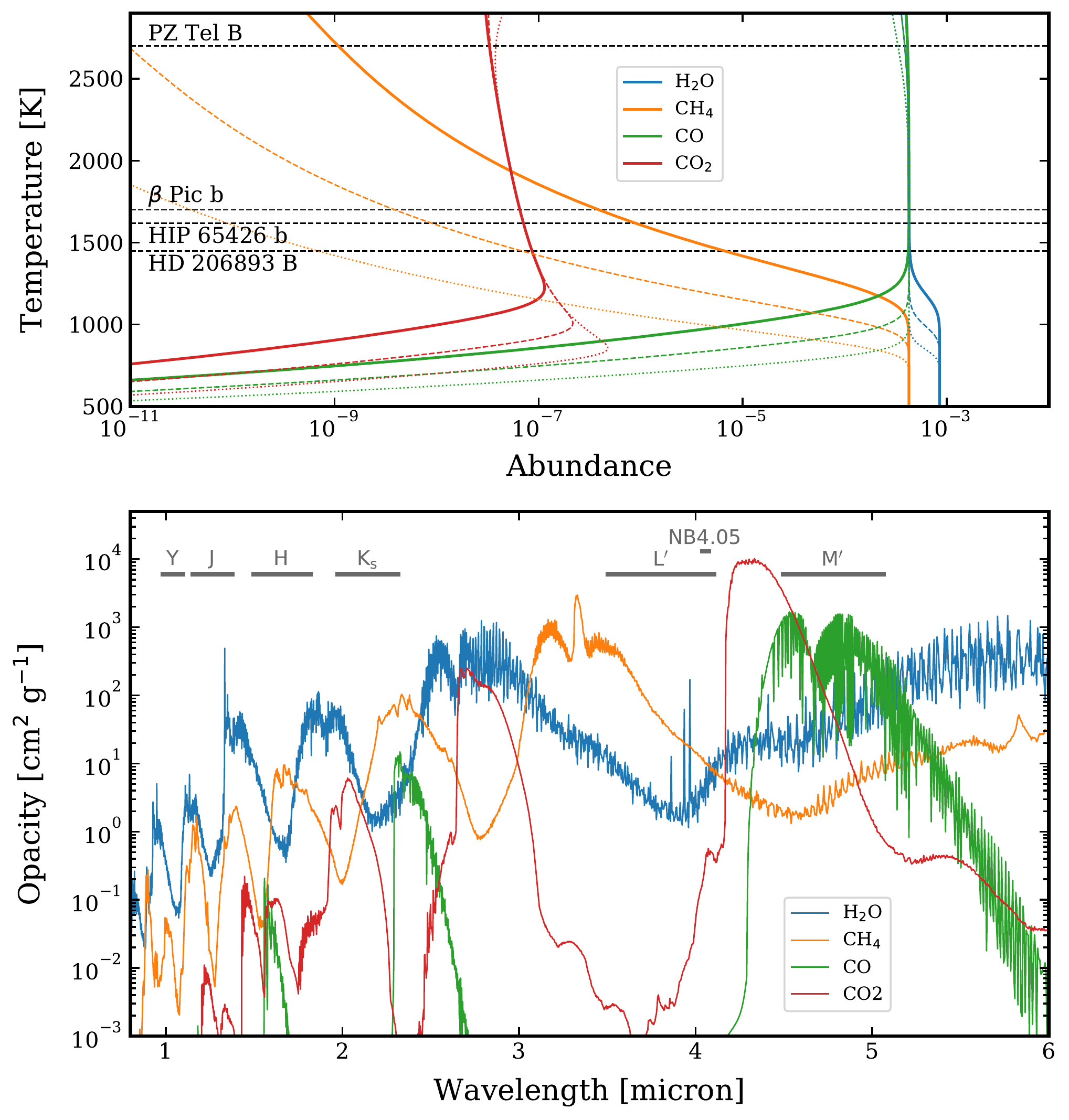}
\caption{\emph{Top panel}: Chemical equilibrium abundances calculated with \texttt{rate} at 1~bar (\emph{solid lines}), 100~mbar (\emph{dashed lines}), and 10~mbar (\emph{dotted lines}) \citep{cubillos2019}. For reference, model-dependent effective temperatures of $\beta$~Pic~b (derived in Sect.~\ref{sec:beta_pic}) and of the rest of the sample \citep{maire2016,delorme2017,cheetham2019} are indicated by horizontally dashed lines. \emph{Bottom panel}: Molecular opacities calculated with \texttt{petitRADTRANS} at 1~bar and 1500~K \citep{molliere2019}. Filter bandpasses are indicated with horizontal bars.\label{fig:molecules}}
\end{figure*}

Before discussing the extracted photometry and colors of our sample, we qualitatively describe the impact of molecular species and clouds on the appearance of gas giant and brown dwarf atmospheres at 3--5~$\mu$m. The photometric characteristics are shaped by the line opacities of the gas and continuum opacity of the cloud condensates. The composition of the gas depends on the temperature and pressure, which is most straightforwardly computed for a set of elemental abundances by assuming that the reactions occur in chemical equilibrium. As an example, we show in the top panel of Fig.~\ref{fig:molecules} how the equilibrium abundances of the most relevant species change with temperature while fixing the pressure. For the same species, we display in the bottom panel of Fig.~\ref{fig:molecules} the molecular opacities at a spectral resolution of 1000.

The spectral window of the $L'$ filter covers the fall and rise of two H$_2$O bands and also part of the CH$_4$ band at 3.3~$\mu$m. However, given the high temperatures ($T_\mathrm{eff} \gtrsim 1300$~K) and low surface gravity of the studied sample, the abundance of CH$_4$ is expected to be low. Specifically, there is evidence that the CO/CH$_4$ chemistry is not in equilibrium in the atmospheres of low-gravity objects \citep[e.g.,][]{konopacky2013,moses2016}. Therefore, only absorption by H$_2$O is expected to affect the $L'$ photometry of our sample. The NB4.05 filter does not cross any significant molecular absorption and is therefore a good tracer of the continuum. The opacity of H$_2$O increases between $L'$ and NB4.05, which is continued across the $M'$ filter. However, the opacity of CO is more than an order of magnitude larger than H$_2$O in the $M'$ range. Therefore, CO will mostly affect the $M'$ photometry, assuming that the CO and H$_2$O abundances are similar. The CO$_2$ fundamental stretching-mode band at 4.2~$\mu$m partially covers the $M'$ band but its equilibrium abundance is about four orders of magnitude smaller than CO in the most optimistic case.


Since the NB4.05 filter is mostly sensitive to continuum emission, both the $L'$~--~NB4.05 and NB4.05~--~$M'$ colors contain some information about the chemical composition of the 
atmosphere. Specifically, the $L'$~--~NB4.05 colors of our sample are expected to be affected by the H$_2$O opacity. The synthetic colors in Fig.~\ref{fig:color_mag2} show that H$_2$O may indeed cause a red color at high temperatures which is further enhanced with the presence of clouds. The color has typical values of only a few tens of a magnitude, except for the low-temperature predictions for which CH$_4$ absorption occurs in $L'$. Similar effects of composition and clouds can be seen at 4--5~$\mu$m. At high temperature, both the predictions of the $L'$~--~$M'$ color (see right panel in Fig.~\ref{fig:color_mag1}) and NB4.05~--~$M'$ color (not shown) indicate blue values in the range of $-0.6$--$0.0$~mag. This occurs due to absorption by CO in $M'$ if the effect of clouds is minimal. With clouds, the AMES-Dusty models predict NB4.05~--~$M'$ colors that are close to gray while the $L'$~--~$M'$ colors show a strong reddening towards lower-mass objects.

\subsection{Photometric characteristics of the individual objects}\label{sec:individual_objects}

\begin{figure*}
\centering
\includegraphics[width=0.9\linewidth]{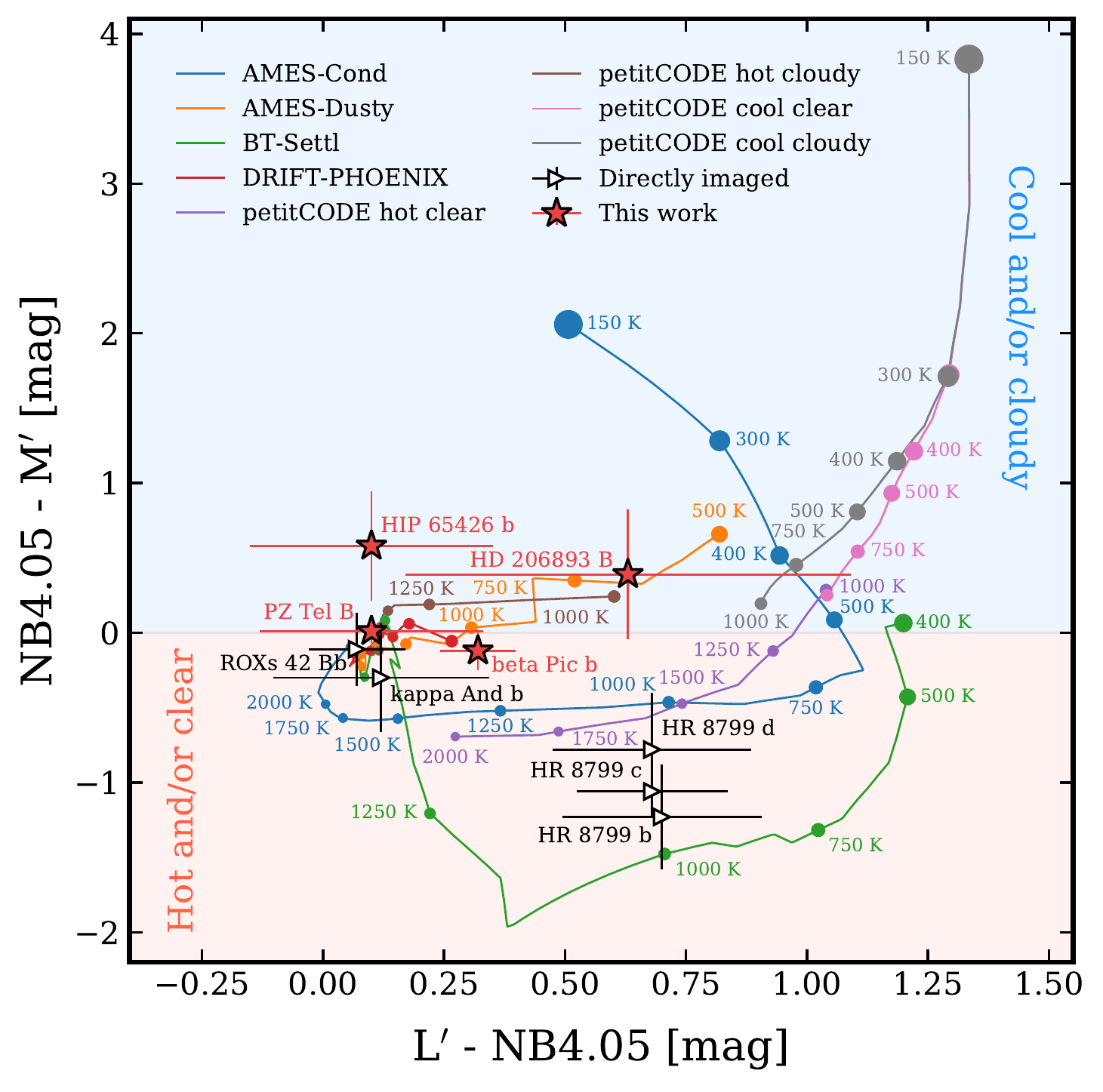}
\caption{Color--color diagram of $L'$~--~NB4.05 vs. NB4.05~--~$M'$. Synthetic colors are calculated from the AMES-Cond ($T_\mathrm{eff} = 150\mathrm{-}3000$~K), AMES-Duty ($T_\mathrm{eff} = 500\mathrm{-}3000$~K), BT-Settl ($T_\mathrm{eff} = 400\mathrm{-}3000$~K), DRIFT-PHOENIX ($T_\mathrm{eff} = 1000\mathrm{-}3000$~K), petitCODE cool ($T_\mathrm{eff} = 150\mathrm{-}1000$~K), and petitCODE hot ($T_\mathrm{eff} = 1000\mathrm{-}2000$~K) atmospheric models (\emph{colored lines}) across the indicated temperature range. The other parameters have been fixed to $\log{g} = 4.0$~dex, $\mathrm{[Fe/H]} = 0.0$~dex, $\mathrm{C/O} = 0.55$ (only petitCODE), and $f_\mathrm{sed} = 2.0$ (only petitCODE cloudy). The \emph{dots} decrease in size towards higher temperatures and correspond to 150, 300, 400, 500, 750, 1000, 1250, 1500, 1750, 2000, and 3000~K. The directly imaged companions for which photometry is available in the three filters are shown for comparison. The \emph{blue} and \emph{red} shaded regions indicate approximately two different NB4.05~--~$M'$ regimes. \label{fig:model_colors}}
\end{figure*}

The magnitudes and colors in the 3--5~$\mu$m range of $\beta$~Pic~b, HIP~65426~b, PZ~Tel~B, and HD~206893~B are presented in Sect.~\ref{sec:fluxes_colors}. Here, we discuss their photometric characteristics in the context of previous studies and the color predictions from atmospheric models. A color--color comparison based on 3--5~$\mu$m photometry of the directly imaged objects and various atmospheric models is therefore shown in Fig.~\ref{fig:model_colors}.

\subsubsection{$\beta$~Pictoris~b}

The apparent magnitude of $\beta$~Pic~b that we derived from the NB4.05 data is $10.98 \pm 0.04$~mag. This value is  consistent with $11.20 \pm 0.23$~mag from \citet{quanz2010} and $11.04 \pm 0.08$~mag from \citet{currie2013}, within the respective uncertainties. In NB3.74, the photometry of the object is $11.25 \pm 0.26$~mag, which implies that the NB3.74~--~$L'$ color is $-0.05 \pm 0.27$~mag \citep[$L' = 11.30 \pm 0.06$~mag;][]{stolker2019}. A (close to) gray color might indeed be expected since the central wavelength at NB3.74 lies in the center of the $L'$ band, which covers a continuous spectral slope (see Fig.~\ref{fig:betapic_spectrum}). In Sect.~\ref{sec:beta_pic}, we constrained the surface gravity of the planet to $\log{g} = 4.17_{-0.13}^{+0.10}$~dex by combining the photometry from $Y$ to $M'$ band and using prior on its mass. This value confirms the low-gravity nature of this object and is consistent with the $\log{g} = 4.18 \pm 0.01$~dex that was derived by \citet{chilcote2017} from the bolometric luminosity and age of the planet.

Figure~\ref{fig:model_colors} shows that the NB4.05~--~$M'$ color of $\beta$~Pic~b ($-0.12 \pm 0.13$~mag) is in line with the predictions by the DRIFT-PHOENIX, AMES-Dusty, BT-Settl, and cloudy petitCODE models (i.e., all cloudy atmospheres). This blue color is likely caused by the CO fundamental band in $M'$ regime. The red $L'$~--~NB4.05 color ($0.32 \pm 0.08$~mag) might be a result of clouds, possibly combined with H$_2$O absorption in $L'$. This color is redder than what is predicted by these same cloudy atmosphere models and is more comparable to a $\sim$1000~K atmosphere. This may indicate that $\beta$~Pic~b has thicker clouds and/or stronger H$_2$O absorption. Such clouds cause an overall reddening of the spectrum, as is the case for the $L'$~--~$M'$ color (see Fig.~\ref{fig:color_mag1}).

The sampled spectra from the MCMC posteriors in Sect.~\ref{sec:beta_pic} point indeed towards a solution with some absorption by H$_2$O in the 3--4~$\mu$m region and CO absorption across the $M'$ band (see Fig.~\ref{fig:betapic_spectrum}). The strength of the absorption features that affect the precise values of the colors will also depend on the elemental abundances of carbon and oxygen. These are included in the combined metallicity parameter of the DRIFT-PHOENIX models which therefore restricts the fit to abundances only relative to solar values.

\subsubsection{HIP~65426~b}

The $L'$ and $M'$ photometry of HIP~65426~b was previously reported by \citet{cheetham2019}. The authors determined a contrast of $8.47 \pm 0.14$~mag and $8.2 \pm 0.4$~mag in $L'$ and $M'$, respectively, and an apparent magnitude of $15.26 \pm 0.15$~mag and $15.1 \pm 0.5$~mag in the two filters. The contrast that we obtained in $L'$ is 0.09~mag larger while the $M'$ contrast is 0.31~mag smaller. A potential difference for the $M'$ data could be related to the PSF template that is used. Specifically, we applied a one-on-one injection of the negative PSF for each image (see Sect.~\ref{sec:error_budget}), thereby accounting for changes in the observing conditions. The derived apparent magnitude is 0.06~mag larger in $L'$ and 0.45~mag smaller in $M'$ compared to the results from \citet{cheetham2019}. Therefore, the flux calibration approach introduced an additional, minor difference in the final magnitudes.

The $L'$~--~NB4.05 color of HIP~65426~b is close to gray ($0.1 \pm 0.25$~mag). Figure~\ref{fig:model_colors} shows that the color is consistent with all model predictions at the approximate $T_\mathrm{eff}$ of the planet ($\sim$1600~K), except for the clear petitCODE model (due to deeper CH$_4$ absorption in the $L'$ band). The $L'$~--~NB4.05 color of HIP~65426~b is less red compared to $\beta$~Pic~b (see also Fig.~\ref{fig:color_mag2}), which may indicate that absorption by H$_2$O in the $L'$ band is dampened more strongly by the atmosphere's dust content. Similarly, the red NB4.05~--~$M'$ color ($0.58 \pm 0.36$~mag) hints at a lack of CO absorption in the $M'$ band due to enhanced cloud densities close to the photosphere. Indeed, the $M'$ flux is relatively large and comparable to a late M dwarf (see Fig.~\ref{fig:color_mag1}). Considering the uncertainty on the 4--5~$\mu$m color, it is most similar (within 1--2$\sigma$) to the synthetic colors from the BT-Settl, DRIFT-PHOENIX, and cloudy petitCODE models.

\subsubsection{PZ~Telescopii~B}

The highly variable observing conditions of the $L'$ dataset resulted in flux variations up to 40\% (see Sect.~\ref{sec:observations}). Although we applied a stringent frame selection to ensure that the science and flux exposures were obtained in comparable conditions, a remaining variation in the stellar flux may have caused a bias in the flux calibration which is difficult to correct or account for in the error budget. The derived contrast in $L'$ of $4.78 \pm 0.21$~mag is 0.37~mag smaller than the value reported  by \citet{beust2016} ($5.15 \pm 0.15$~mag), which could be related to a difference in frame selection. The contrast from \citet{beust2016} was calibrated by \citet{maire2016} to an apparent magnitude of $11.05 \pm 0.18$~mag. Therefore, our final magnitude is only 0.01~mag smaller due to our different approach with the flux calibration.

The $L'$~--~NB4.05 color of PZ~Tel~B is slightly red ($0.10 \pm 0.23$~mag), but also consistent with being gray or blue, while the NB4.05~--~$M'$ is gray ($0.01 \pm 0.08$~mag). At 2700~K, the $L'$~--~NB4.05 and NB4.05~--~$M'$ colors are expected to be red and blue (by a tenth of a magnitude), respectively, due to absorption by H$_2$O and CO. Figure~\ref{fig:model_colors} shows that all model predictions converge at the highest temperatures because cloud condensation will not occur. Consequently, the differences in the sample of high- and low-gravity objects is more limited (see Figs.~\ref{fig:color_mag1} and \ref{fig:color_mag2}). The color--color diagram of Fig.~\ref{fig:model_colors} shows a clustering of synthetic colors close to gray, which approximately coincides with the colors of PZ~Tel~B (see also Fig.~\ref{fig:color_color}). The slight reddening (although not significant) of the $L'$~--~$M'$ color could be caused by a thin layer of small cloud particles which may have formed at higher altitudes than predicted by the considered models.

\subsubsection{HD~206893~B}

The low-mass companion of HD~206893 was previously identified as an extremely red, late L dwarf \citep{milli2017,delorme2017}. The precise nature of the object is not clear because of the uncertain age of the system, yielding mass estimates in the range of 12--50~$M_\mathrm{Jup}$ \citep{delorme2017}. The $L'$ brightness of HD~206893~B that we derived in the present study is $13.80 \pm 0.31$~mag, which implies that the object is 0.37~mag fainter compared to results from \citet{milli2017} (a similar stellar magnitude was used for the calibration). However, the values are comparable within their uncertainties, which are relatively large due to residual speckle noise at the small separation of HD~206893~B.

With a lower $L'$ brightness and less red color, the object's photometric characteristics overlap with the low-gravity objects in the $H$~--~$L'$ versus M$_{L'}$ diagram (see left panel in Fig.~\ref{fig:color_mag1}). The high absolute flux in $M'$, on the other hand, results in a very red $L'$~--~$M'$ color, which deviates by $\sim$2$\sigma$ from the field dwarfs (see Fig.~\ref{fig:color_mag1}). The $L'$~--~NB4.05 color of HD~206893B is redder than both $\beta$~Pic~b and HIP~65426~b, although the error bar is large. At an approximate temperature of 1450~K, methane absorption is expected to occur in the $L'$ band in cases where abundances are in chemical equilibrium (see AMES-Cond and clear petitCODE predictions in Fig.~\ref{fig:model_colors}). However, in a low-gravity environment, quenching of CH$_4$ and CO is expected to occur at the considered temperature, which would increase the $L'$ flux. Furthermore, the NB4.05~--~$M'$ color is red, but is approximately in line with predictions by the AMES-Dusty, petitCODE cloudy, and DRIFT-PHOENIX models at those temperatures. The combined photometric characteristics may hint at an extended amount of dust in the atmosphere of HD~206893~B or the presence of circumplanetary material which partially reprocesses its photospheric emission.

\subsection{Upcoming opportunities for MIR imaging observations of planetary and substellar companions}\label{sec:upcoming_opportunities}

As discussed above, adding MIR information to the SEDs of low-mass objects puts additional constraints on their atmospheric properties. In addition, the MIR allows in principle to search for cooler -- and hence older or lower mass -- objects than possible at shorter wavelengths. For cooler planets, there will be an increasing amount of information available from photometry in the 3--5~$\mu$m regime because the peak flux shifts towards the MIR and molecular absorption features increase in strength. Color effects between the $L'$, NB4.05, and $M'$ fluxes are therefore expected to become stronger towards lower temperatures (see Fig.~\ref{fig:model_colors}).

From the ground, we expect the ERIS instrument for the VLT \citep{davies2018} to be an important step forward and to open up new discovery space in both new detections and better characterization potential. The ERIS instrument will sit behind the adaptive secondary mirror of the Adaptive Optics Facility and provide better AO performance than NACO. In combination with an overall reduced thermal background (due to fewer warm surfaces in the light-path compared to NACO) and state-of-the-art coronagraphs, ERIS will be significantly more sensitive. Furthermore, ERIS will have spectroscopic capabilities in the 3--5~$\mu$m range meaning that photometry can be complemented with medium-resolution spectra. 

In terms of sensitivity and wavelength coverage in the MIR, \emph{JWST} is expected to set new benchmarks for years to come. Depending on the final contrast performance, some of the known low-mass and planetary companions may be accessible to \emph{JWST} instrumentation and their atmospheres can be probed for a variety of constituents including nitrogen- and phosphorus-bearing species such as NH$_3$ and PH$_3$ \citep{danielski2018}. In addition, the broad spectral coverage will help  determine the bolometric luminosity of these young objects, thereby putting additional constraints on evolutionary models \citep{marleau2019}. Motivated by the expected significant increase in sensitivity, new evolutionary models were recently introduced \citep{linder2019} to make predictions for objects down to masses comparable to those of the ice giants in the Solar System. Whether or not \emph{JWST} will indeed find such objects around young nearby stars remains to be seen.

Finally, as one of the first-generation instruments, the 39m Extremely Large Telescope will feature the Mid-IR ELT Imager and Spectrograph \citep[METIS;][]{brandl2018}. While not being able to compete with \emph{JWST} in terms of sensitivity, the unparalleled spatial resolution of METIS and its ability to point to the brightest stars in the sky will allow METIS to search for and potentially image small, terrestrial exoplanets around stars in the immediate vicinity of the Sun \citep{quanz2015a}. The first experiment demonstrating the general feasibility of such observations was recently carried out in the context of the NEAR campaign with the upgraded VISIR instrument at the VLT \citep{kasper2017}. 

\begin{acknowledgements}

We thank Greta Guidi, Gabriele Cugno, and Janis Hagelberg for insightful discussions. We also thank the referee whose constructive comments improved the quality of this manuscript. T.S. acknowledges the support from the ETH Zurich Postdoctoral Fellowship Program. Part of this work has been carried out within the framework of the National Centre of Competence in Research PlanetS supported by the Swiss National Science Foundation. S.P.Q. acknowledges the financial support of the SNSF. P.M. and K.T. acknowledge support from the European Research Council under the European Union's Horizon 2020 research and innovation programme (grant agreement No. 694513 and 832428, and 679633, respectively). This research has made use of the SVO Filter Profile Service\footnote{\url{http://svo2.cab.inta-csic.es/theory/fps/}} supported from the Spanish MINECO through grant AYA2017-84089.

\end{acknowledgements}

\bibliographystyle{aa}
\bibliography{references}

\begin{appendix}

\section{Posterior probability distributions}\label{sec:appendix_posteriors}

\begin{figure*}
\centering
\includegraphics[width=\linewidth]{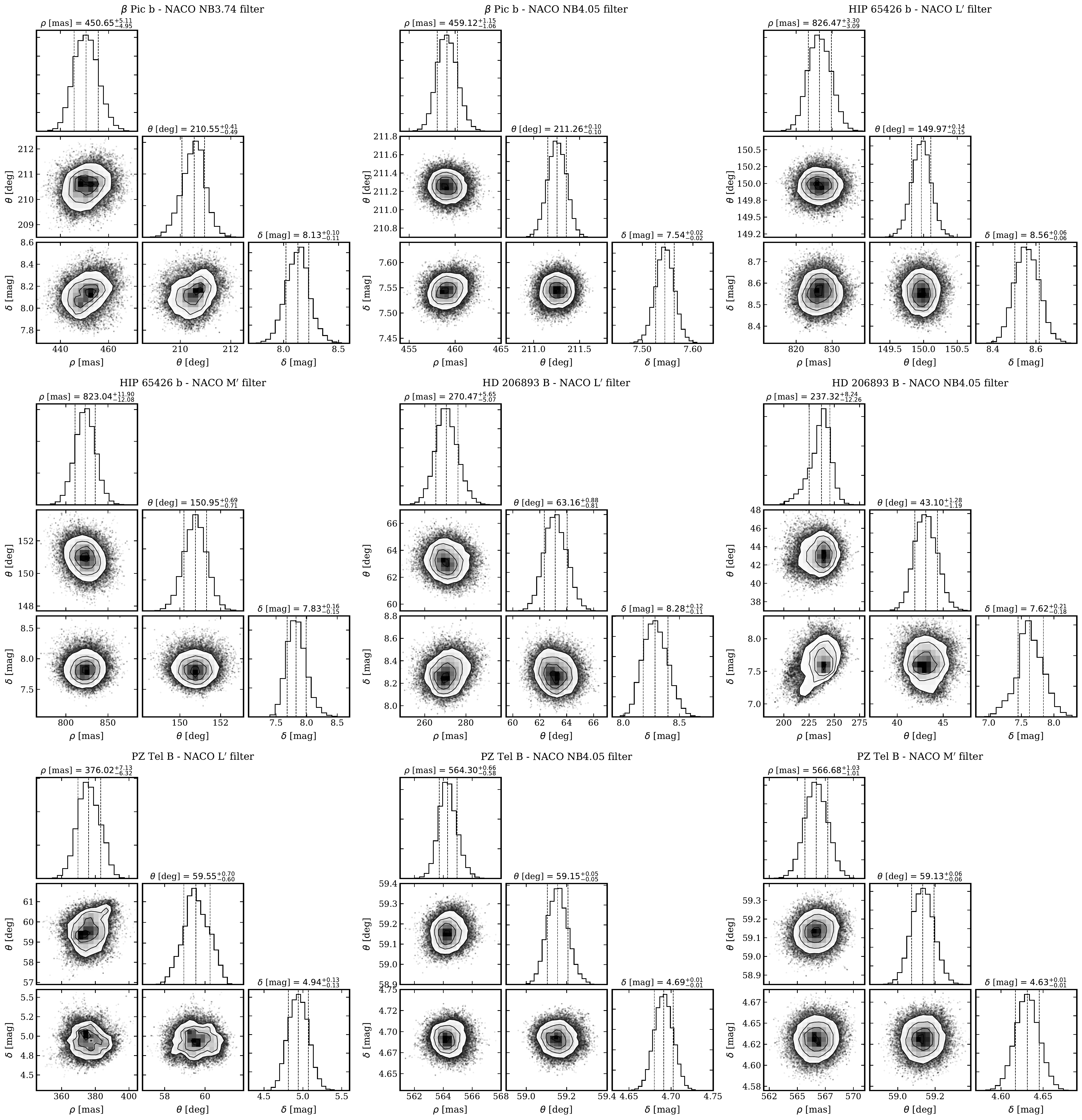}
\caption{Posterior distributions of the separation, $\rho$, position angle, $\theta$, and planet-to-star flux contrast, $\delta$, for the remaining companions and filters. The companion name and filter is provided above each panel. Additional information can be found in the caption of Fig.~\ref{fig:mcmc1}.\label{fig:mcmc2}}
\end{figure*}

In addition to Fig.~\ref{fig:mcmc1}, we present the posterior distributions for the remaining companions and filters in Fig.~\ref{fig:mcmc2}.

\section{Astrometry of the companions}\label{sec:appendix_astrometry}

The astrometry of the companions is listed in Table~\ref{table:astrometry}. The overview includes the retrieved separation and position angle (measured east of north) from the MCMC analysis described in Sect.~\ref{sec:relative_calibration} and related posterior distributions presented in Figs.~\ref{fig:mcmc1} and \ref{fig:mcmc2}. The measurement bias and related uncertainties are also listed, as described in Sect.~\ref{sec:error_budget} and illustrated in Fig.~\ref{fig:error}.

\begin{sidewaystable*}
\caption{Astrometry and error budget.}
\label{table:astrometry}
\centering
\bgroup
\def\arraystretch{1.25}
\begin{tabular}{L{2cm} L{1.4cm} C{2cm} C{2.1cm} C{1.9cm} C{2cm} C{1.8cm} C{2cm} C{2cm}}
\hline\hline
Target & Filter & Separation MCMC\tablefootmark{a} & Separation bias\tablefootmark{a} & P.A. MCMC\tablefootmark{a} & P.A. \newline bias\tablefootmark{a} & True north correction & Final separation\tablefootmark{b} & Final P.A.\tablefootmark{c} \\
 & & (mas) & (mas) & (deg) & (deg) & (deg) & (mas) & (deg) \\
\hline
$\beta$ Pic b & NB3.74 & $450.65 \pm 5.11$ & $0.66 \pm 12.58$ & $210.55 \pm 0.49$ & $-0.02 \pm 1.30$ & $-0.40 \pm 0.10$ & $451.30  \pm 13.58$ & $210.13  \pm 1.44$ \\
$\beta$ Pic b & NB4.05 & $459.12 \pm 1.15$ & $0.03 \pm 2.58$ & $211.26 \pm 0.10$ & $-0.01 \pm 0.20$ & $-0.40 \pm 0.10$ & $459.15  \pm 2.83$ & $210.84  \pm 0.46$ \\
HIP 65426 b & $L'$ & $826.47 \pm 3.30$ & $0.79 \pm 7.57$ & $149.97 \pm 0.15$ & $-0.00 \pm 0.35$ & $-0.40 \pm 0.10$ & $827.26  \pm 8.26$ & $149.56  \pm 0.55$ \\
HIP 65426 b & NB4.05 & $825.33 \pm 9.93$ & $-1.56 \pm 17.10$ & $150.13 \pm 0.49$ & $-0.08 \pm 1.06$ & $-0.40 \pm 0.10$ & $823.77  \pm 19.77$ & $149.65  \pm 1.23$ \\
HIP 65426 b & $M'$ & $823.04 \pm 12.08$ & $1.79 \pm 18.75$ & $150.95 \pm 0.71$ & $0.01 \pm 1.09$ & $-0.40 \pm 0.10$ & $824.83  \pm 22.30$ & $150.56  \pm 1.36$ \\
PZ Tel B & $L'$ & $376.02 \pm 7.13$ & $-10.01 \pm 10.74$ & $59.55 \pm 0.70$ & $0.01 \pm 0.47$ & $-0.40 \pm 0.10$ & $366.01  \pm 12.89$ & $59.16  \pm 0.94$ \\
PZ Tel B & NB4.05 & $564.30 \pm 0.66$ & $-0.08 \pm 1.43$ & $59.15 \pm 0.05$ & $0.00 \pm 0.11$ & $-0.40 \pm 0.10$ & $564.22  \pm 1.57$ & $58.76  \pm 0.42$ \\
PZ Tel B & $M'$ & $566.68 \pm 1.03$ & $0.15 \pm 1.78$ & $59.13 \pm 0.06$ & $-0.01 \pm 0.15$ & $-0.40 \pm 0.10$ & $566.84  \pm 2.06$ & $58.72  \pm 0.43$ \\
HD 206893 B & $L'$ & $270.47 \pm 5.65$ & $-0.95 \pm 10.76$ & $63.16 \pm 0.88$ & $0.00 \pm 1.93$ & $-0.40 \pm 0.10$ & $269.53  \pm 12.15$ & $62.76  \pm 2.16$ \\
HD 206893 B & NB4.05 & $237.32 \pm 12.26$ & $1.80 \pm 12.55$ & $43.10 \pm 1.28$ & $-0.17 \pm 1.71$ & $-0.40 \pm 0.10$ & $239.12  \pm 17.55$ & $42.53  \pm 2.17$ \\
HD 206893 B & $M'$ & $245.33 \pm 10.18$ & $1.18 \pm 18.76$ & $43.09 \pm 1.00$ & $0.11 \pm 1.96$ & $-0.40 \pm 0.10$ & $246.51  \pm 21.34$ & $42.80  \pm 2.24$ \\
\hline
\end{tabular}
\egroup
\tablefoot{\\
\tablefoottext{a}{For asymmetric uncertainties, we conservatively use the largest of the two values that were derived from both the statistical and systematic error analysis (see Figs.~\ref{fig:mcmc1}, \ref{fig:error}, and \ref{fig:mcmc2}).}\\
\tablefoottext{b}{The final separation is calculated by adding the bias offset and combining the two error components in quadrature.}\\
\tablefoottext{c}{The final position angle (P.A.) is calculated by adding the bias and assumed true north offset, and combining the three error components in quadrature.}\\
}
\end{sidewaystable*}

\section{Color corrections for WISE and NACO}\label{sec:appendix_calibration}

\begin{figure*}
\centering
\includegraphics[width=\linewidth]{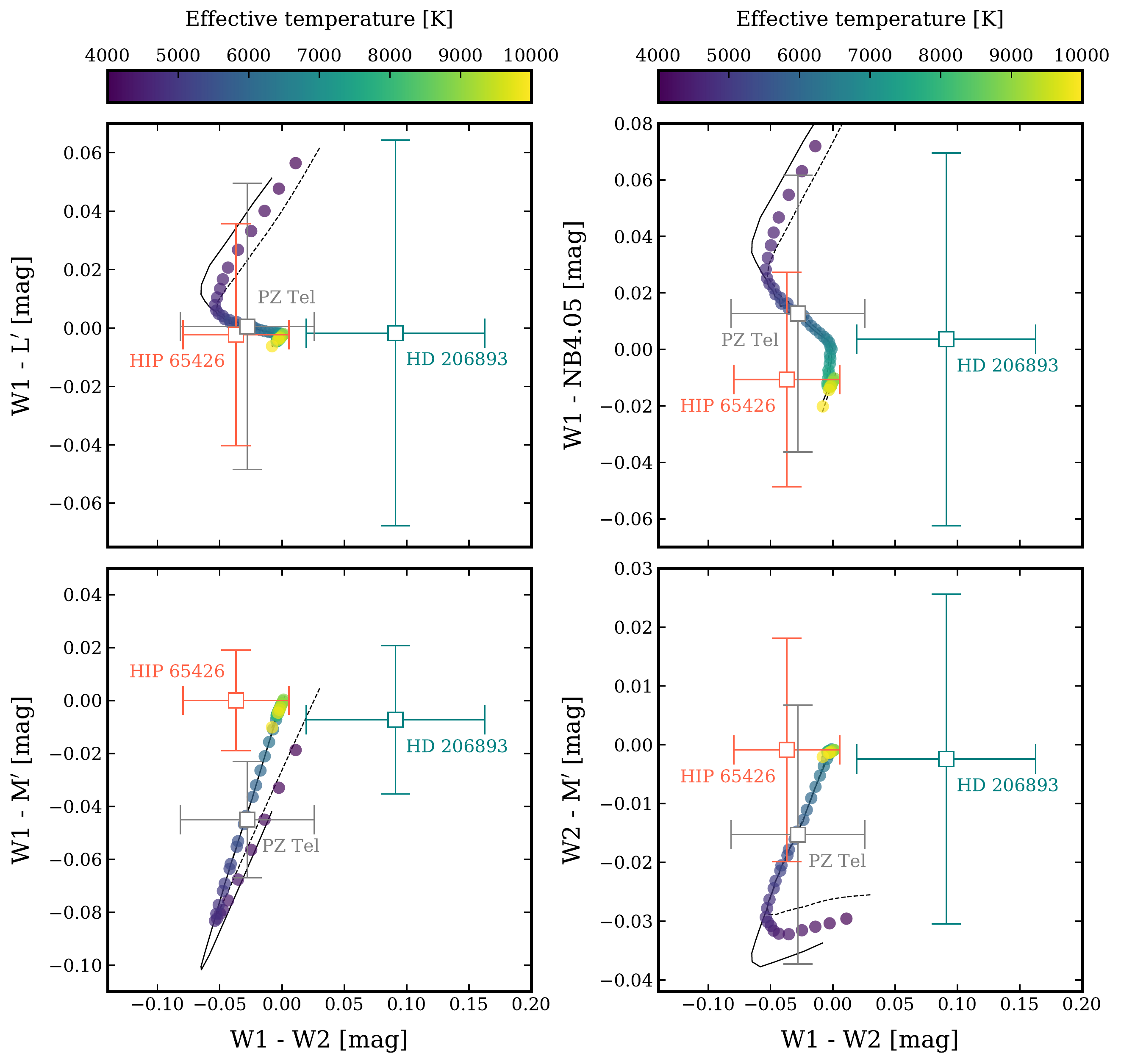}
\caption{Color--color diagrams of $W1$~--~$W2$ vs. $W1$~--~$L'$ (\emph{top left panel}), $W1$~--~NB4.05 (\emph{top right panel}), $W1$~--~$M'$ (\emph{bottom left panel}), and $W2$~--~$M'$ (\emph{bottom right panel}). The synthetic photometry (\emph{colored dots}) is calculated for 50 logarithmically spaced temperatures between 4000~ and 10000~K and $\log{g} = 4.5$~dex. For comparison, the same calculation was done for $\log{g} = 4.0$~dex (\emph{black solid line}) and $\log{g} = 5.0$~dex (\emph{black dashed line}). The individual data points show the $W1$~--~$W2$ color and uncertainty of HIP~65426, PZ~Tel, and HD~206893 along the horizontal axis (see Table~\ref{table:targets}). For the vertical axis, the color was computed from the BT-NextGen models at an effective temperature of 8840~K \citep[HIP~65426;][]{chauvin2017}, 5665~K \citep[PZ~Tel;][]{schmidt2014}, and 6500~K \citep[HD 206893;][]{delorme2017} with the error bars showing the uncertainty on either the $W1$ or $W2$ photometry.\label{fig:wise_naco}}
\end{figure*}

The photometry of HIP~65426, PZ~Tel, and HD~206893 was calibrated from the available WISE photometry. We computed colors of the WISE and NACO filters from the BT-NextGen atmospheric models \citep{allard2012} in order to determine the color corrections that are required as a function of effective temperature. The color--color diagrams in Fig.~\ref{fig:wise_naco} show the comparison of the synthetic colors with the $W1$~--~$W2$ colors of our sample.

\end{appendix}

\end{document}